\documentclass[prd,aps]{revtex4}
\usepackage{color}
\usepackage{mathtools}
\usepackage{array}
\usepackage{tabularx}
\usepackage{diagbox}
\usepackage{tikz}
\usetikzlibrary{decorations.pathmorphing}
\usepackage{bm}
\usepackage{amsmath,bm,amssymb,amsfonts,dcolumn,color,graphicx,latexsym,epsfig}
\usepackage{bbold}
\usepackage{rsfso}
\usepackage{microtype}
\usepackage{setspace} 
\usepackage{array, booktabs}
\usepackage[utf8]{inputenc}
\UseRawInputEncoding
\usepackage{hyperref}
\hypersetup{
colorlinks=true,
linkcolor=blue,
filecolor=magenta,      
urlcolor=magenta,
citecolor=magenta,
}
\usepackage{multirow}
\usepackage{natbib}
\usepackage{float}
\usepackage{booktabs}
\usepackage{pifont}
\usepackage{mathrsfs}
\usepackage{caption}
\usepackage{caption, threeparttable}

\begin{document}

\title{Quasinormal spectra of a wormhole family: overtone features 
and a parameter-controlled redshift}
\author{Abhisek Barman Maji}
\email[Email address: ]{rajyavisek027@kgpian.iitkgp.ac.in}
\affiliation{Department of Physics, Indian Institute of Technology, Kharagpur 721 302, India}

\author{Sayan Kar}
\email[Email address: ]{sayan@phy.iitkgp.ac.in}
\affiliation{Department of Physics, Indian Institute of Technology, Kharagpur 721 302, India}

\begin{abstract}
\noindent Though investigated extensively in the past, we take a detailed relook 
at the study of quasinormal modes (QNM) in a known family of wormholes
(ultrastatic ($g_{00}=-1$), as well as spacetimes with different, non-constant $g_{00}$),
which includes the familiar Bronnikov-Ellis spacetime as a special case.
Our focus here is to go beyond the fundamental mode and 
obtain some of the QNM overtones using a suitable numerical scheme. Scalar and axial gravitational QNMs including two or three
overtones are obtained for the family of ultrastatic geometries and their parameter dependencies are shown
explicitly. Further, we comment on 
how (a) the overtones may influence the time-domain profile in a perturbation
and (b) in what sense, the use of overtones may help in distinguishing between different geometries within the family.

\noindent Finally, we show how an effect somewhat similar to the so-called `environment induced redshift' of QNMs, introduced recently (Phys. Rev. D \textbf{111}, 064026 (2025)), may be obtained for spacetimes with non-constant $g_{00}$,
via an appropriate tuning of available metric parameters which 
systematically modify the
shapes of the effective potentials arising in the perturbation
equations. 

\noindent 

\noindent 
\end{abstract}


\maketitle

\section{Introduction and summary} \label{introduction}

\noindent The notion of a Lorentzian wormhole spacetime appeared first in the well-known work of Einstein and Rosen \cite{einstein1935particle}, popularly referred to as the Einstein-Rosen bridge spacetime. However, it was Wheeler's `geons' \cite{Wheeler1955Geons} in the 1950s and 1960s which highlighted wormhole topology of the electric field lines as a model of `charge without charge' and also introduced the term `wormhole' \cite{Misner1957Classical} in the literature. Since then, there have been important landmarks in `wormholery'. The early 1970s saw the appearance of the first `exact wormhole solution' in General Relativity coupled to matter, {\em albeit} with a somewhat odd (negative energy density) matter field. This rather simple geometry is now referred to as Bronnikov-Ellis
geometry \cite{bronnikov1973scalar, ellis1973ether}. In the late 1980s, Euclidean wormholes were very popular among Euclidean quantum gravity proponents at that time. Around the same time, Morris and Thorne \cite{morris1988wormholes} and
Morris-Thorne-Yurtsever \cite{morris1988wormholes1} came up with a couple of seminal papers which re-invented the wormhole idea in a Lorentzian context by bringing in aspects of energy condition violations, traversability and the
exotic wormhole time-machine model. The Morris-Thorne paradigm
was hotly pursued by many--most notably by Visser \cite{visser1995lorentzian} who came up with numerous constructions (including the cut-and-paste method, surgery ) of Lorentzian wormholes. 

\noindent The entire scenario was largely theoretical, with very little effort being given to even suggest possible observational consequences. A large part of the efforts focused on how to avoid energy condition violations \cite{Visser2003Traversable, Maeda2008Static}, and modified/alternate theories of gravity \cite{Bhawal1992Lorentzian, Kanti2012Stable, Hochberg1990Lorentzian, Agnese1995Wormholes, Shaikh2016Wormholes, Kumar2024, Radhakrishnan2024, Dehghani2009, Sadatian2024, Loewer2024, Battista2024} seemed to be a sort of way out. It was noticed that in a modified theory of gravity,
the purely geometric violation of the timelike or null convergence condition does not necessarily imply a violation of the weak, null or dominant energy conditions. Hence, energy conditions, postulated as physical requirements on matter, may hold even though the convergence conditions may be violated, as required for a wormhole to exist. Thus,
wormholes seemed more viable in various modified theories of gravity. 

\noindent Given the above fact that wormholes could be around if any modified theory of gravity is acceptable, even in a restricted domain of the theory parameters, it is obvious that one must find observational signatures. Gravitational lensing of wormholes \cite{Cramer1995, DEY2008GRAVITATIONAL, Shaikh2019Strong, Godani2021Gravitational, Nandi2006Gravitational, EIROA2001, Izmailov2019, Mitra2024, Tsukamoto2017} was one of the first among the observationally relevant works pursued by researchers. It still continues as an important possible signature. 

\noindent Over the last decade or so, astronomy and astrophysics has reached a different level of
observational relevance through the emergence of a large 
amount of data and results in gravitational wave astronomy 
(LIGO-VIRGO-KAGRA collaboration \cite{PhysRevX.13.041039}) and infrared imaging of astrophysical objects (the Event Horizon Telescope (EHT) \cite{M872019, EventHorizonTelescopeCollaboration2022}). The results obtained tally well with General Relativity predictions, leaving little room for modified gravity. However, there does remain a small but definite possibility of accommodating a modified gravity solution or a sort of exoticity within GR (eg. the Bronnikov-Ellis geometry).  It is this possibility which has driven some researchers (including us) to construct models, do calculations 
with the aim of looking for signatures of wormholes in available data. With more observed events and data in the future, one can always hope for further confirmation of GR or, perhaps, some acceptance of a modification of GR, of course, within a specific domain of parameter values.

\noindent One of the central observationally important arenas in GW astrophysics, which has been flooded, in the last decade,  by a huge amount of theoretical work by various groups, is the study of quasinormal modes \cite{VISHVESHWARA1970} in perturbation theory (see \cite{bhpqnm} for a very recent review). Scalar and gravitational perturbations have been worked out for diverse black hole \cite{berti2025ringdown, Berti2009, Cardoso2001Quasinormal, Nollert1999, Konoplya2002, konoplya2003gravitational,rincon2020greybody,  BlzquezSalcedo2018Radial} and wormhole \cite{Bueno2018Echoes, Konoplya2005Ringing, kim2008wormhole, alfaro2024quasinormal, Jirn2025} spacetimes in GR as well as in modified gravity theories. Any such study on a spacetime not among the conventional solutions of GR (eg. Schwarzschild or Kerr) is referred to as a study on a `mimicker'. More specifically, wormholes constitute a
prominent family among such `black hole mimickers' \cite{Abdikamalov2019, Konoplya2019Stable, Urbano2019, Narzilloev2020, Olivares2020, Shaikh2022Shadows, Shao2021, Siemonsen2024, Pani2010Gravitational, Cardoso2019Testing, Konoplya2016Wormholes, Glampedakis2018, DeSimone2025}. 
Collectively, such mimickers are also referred to as `exotic compact objects'. The main idea here is that these spacetimes could be among possible stable end-states in a post-merger scenario \cite{SanchisGual2019, Cardoso2016Gravitational, Krishnendu2017}. Whether this is a feasible option at all, will be evident as more observational results emerge.  

\noindent Thus, with the above-mentioned outlook and motivation, here we choose to study quasinormal modes
(scalar and axial gravitational) of a family of wormholes which are generalisations of the well-known Bronnikov-Ellis spacetimes. It must be stated that much has been done with regard to quasinormal modes for these spacetimes in the recent past. So, it's worth asking-- what is it that we intend to focus on, which could be new and interesting?

\noindent Our two-fold aim, in this work, may be stated precisely as follows.

$\bullet$ Most of the work on QNMs of the family of spacetimes studied here is focused on the fundamental QNMs, which surely have the largest contribution (amplitude-wise) in a time domain profile. Despite this, 
with possible detection \cite{capano2023multimode} (though disputed \cite{cotesta2022analysis}) of overtones and the possible detectability of overtones in LISA \cite{oshita2023slowly}, 
it is always important to track down the contributions
 of the so-called `overtones'. This, indeed, has been a line of thought of late, with varied motivations. Researchers were keen on finding overtones from the onset of QNM analysis. Chandrasekhar and Detweiler \cite{chandrasekhar1975} computed the fundamental mode along with a couple of overtones for the Schwarzschild metric. Later, the QNM spectra with overtones for RN \cite{ferrari1984new}, Kerr \cite{Detweiler1980,leaver1985}, Kerr-Newman \cite{Berti2005}, regular blackhole \cite{Konoplya2022, Konoplya2024Infinite, 2402.15085}, parametrized black hole \cite{Hirano2024} metrics were computed. The authors of \cite{kim2008wormhole,batic2024unified,alfaro2024quasinormal,valos2022, Jirn2025, Gonzlez2022, Konoplya2025Transition} computed QNM overtones for various wormhole geometries.
 
 \noindent {\em Our aim here is to first obtain these overtones using a numerical scheme which allows us to calculate as many overtones as possible. Once we have the overtones for the classes of geometries we consider here, we use them to figure out how they may affect the approach to stability for a perturbed spacetime.}

$\bullet$ It has been found (theoretically) that black hole quasinormal modes, when obtained for a scenario where the black hole is in an environment (e.g. a halo), exhibit a `redshift' w.r.t. the increase of the environment parameter \cite{Barausse2014Can, pezzella2024quasinormal, Liu2025, Chen2023}. 

\noindent {\em We show here, for the QNMs and overtones found by us for the wormhole family, that there is indeed a `redshift' when we modify the so-called effective potential by tuning a metric parameter appropriately.}

\noindent Our  article, which is largely a sequel to \cite{dutta2020revisiting}, is organised as follows. In the next section, we provide a summary of the wormhole family we study in this article. In Section \ref{result}, we summarise the numerical results obtained with explicit plots. Section \ref{Inference} discusses the inferences which may be drawn from our studies on the QNMs and overtones. The above-mentioned `redshift' effect is discussed extensively in Section \ref{Redshift}. Finally, in Section \ref{conclusion}, we conclude with remarks on possible future work. Appendix \ref{method} quickly outlines the methods adopted in finding the QNMs--more specifically, which method is better suited for finding overtones is emphasised and discussed in more detail here.  The appendix \ref{Table} contains the detailed numerical values of the QNMs, thereby supporting the plots shown in Section \ref{result}.

\section{A brief review on the family of wormhole geometries} \label{geometry}

\noindent We begin with a brief review of Morris-Thorne wormholes  \cite{morris1988wormholes}(1988) and the theoretical framework for constructing such wormholes within GR. The line element corresponding to a generic,  spherically symmetric, static spacetime is given as
\begin{equation}
    ds^2 = - e^{2\Psi(r)} dt^2 + \frac{dr^2}{1-\frac{b(r)}{r}} + r^2 d\theta^2 + r^2 sin^2\theta d\varphi^2 \label{geo1}
\end{equation}
where, $\Psi(r)$, in wormhole parlance is named the redshift function and $b(r)$, the shape function. Morris--Thorne defined a Lorentzian
traversable wormhole via a set of conditions on $\Psi(r)$ and $b(r)$.
These were: (a) $e^{2\Psi(r)}$ never zero (no horizon), (b) $\frac{b(r)}{r} \leq 1$
with $b(r=b_0) = b_0$ (shape and throat) and (c) $\frac{b(r)}{r} \rightarrow 0$ as $r\rightarrow\infty$ (asymptotic flatness). 

 \noindent Much before the Morris-Thorne paper, in 1973, Bronnikov \cite{bronnikov1973scalar} and  Ellis \cite{ellis1973ether} independently constructed a static, spherically symmetric and horizon-less wormhole solution to the Einstein equations with a minimally coupled phantom scalar field (scalar field with negative kinetic energy). This solution, commonly referred to as Bronnikov--Ellis wormhole, is the simplest Morris-Thorne wormhole with $\Psi(r)=0$ and $b(r)=\frac{b_0^2}{r}$ and is represented by the line element
\begin{equation}
    ds^2 = - dt^2 + \frac{dr^2}{1-\frac{b_0^2}{r^2}} + r^2 d\theta^2 + r^2 sin^2\theta d\varphi^2 \label{geo2}
\end{equation}
One may also rewrite the line element using the proper radial distance,
i.e $r^2= b_0^2+l^2$ (where $-\infty \leq  l  \leq \infty$). In this
representation, we have
\begin{equation}
    ds^2 = -dt^2+ dl^2 + \left (l^2 + b_0^2\right ) d\Omega_2^2
\end{equation}
We may visualise the above line element as representing two asymptotically flat regions ($l\rightarrow \pm \infty$) connected by
a throat/bridge of radius $b_0$. Note that the radial coordinate $r$ runs from $b_0$ to $\infty$ whereas $l$ runs from $-\infty$ to $\infty$ with
$l=0$ corresponding to $r=b_0$. 

\noindent We will now present generalisations of the
above-mentioned simple geometry. The first one will be an ultrastatic spacetime
(i.e. with $g_{00}=-1$) with a different shape function, whereas the other two will have non-constant redshift functions. 

\subsection{The ultra-static wormhole family}

\noindent The Bronnikov-Ellis solution may be viewed as a special case of a more general traversable wormhole solution. Such a generalised Bronnikov--Ellis wormhole was proposed in \cite{kar1995resonances}. Here, $b(r)$ is a two-parameter function ($n$ and $b_0$). When $n=2$, we recover the original Bronnikov--Ellis geometry. The metric of such a generalised Bronnikov--Ellis wormhole \cite{kar1995resonances,dutta2020revisiting} is
\begin{equation}
    ds^2 = - dt^2 +  \frac{dr^2}{1-\frac{b(r)}{r}}+ r^2 d\theta^2 + r^2 sin^2\theta d\varphi^2 \label{geo3}
\end{equation}
\begin{equation*}
    b(r)=r-r^{3-2 n} \left(r^n-b_0^n\right){}^{2-\frac{2}{n}}
\end{equation*}
Alternatively, the line element can be expressed as
\begin{equation}
    ds^2 = - dt^2 +  dl^2+ (l^n+b_0^n)^\frac{2}{n}( d\theta^2 +  sin^2\theta d\varphi^2) \label{geo4}
\end{equation}
Here, $l$  $(\in [-\infty,\infty])$ is the tortoise coordinate 
\begin{align*}
    dl^2 =\frac{dr^2}{1-\frac{b(r)}{r}}
     \implies r(l)=(l^n+b_0^n)^\frac{1}{n}
\end{align*}
We will consider only the even values of the parameter $n$ so that $r(l)$ remains smooth and non-zero everywhere in the domain of $l$.

\subsection{Bronnikov--Ellis wormhole with asymptotic mass} \label{Ellis_asymp_mass}

\noindent Till now, we have discussed the ultrastatic, symmetric and massless Bronnikov--Ellis geometry and its generalisation. However, the authors of \cite{bronnikov1973scalar}, \cite{ellis1973ether} (see also \cite{blazquez2018scalar}) have proposed a metric for a massive, static, Bronnikov--Ellis wormhole for which the line element is of the following form:
\begin{equation}
    ds^2 = -e^{f(l)}dt^2 + e^{-f(l)}[dl^2+(l^2+b_0^2)d\Omega^2]
\end{equation}
with
\begin{equation*}
    f(l)=\frac{C}{b_0}\left(\arctan\left(\frac{l}{b_0}\right)-\frac{\pi}{2}\right)
\end{equation*}
In the asymptotic limit $C=2M$, where $M$ is the mass extracted in the asymptotically flat region. We may extend this geometry further 
by incorporating the $r(l)$ discussed in the previous
subsection. This leads to the new line element given as,
\begin{equation}
    ds^2 = -e^{f(l)}dt^2 + e^{-f(l)}[dl^2+(l^n+b_0^n)^\frac{2}{n}d\Omega^2] \label{AsymM_Ellis}
\end{equation}
For $M=0$, we recover the symmetric generalised Bronnikov--Ellis wormhole, while any non-zero value of $M$ will introduce asymmetry in the wormhole shape. The QNMs for the $n=2$, $M\neq 0$ geometry will be discussed in Section \ref{AsymM}.

\noindent The energy conditions in general relativity serve as criteria to ensure the physical viability of solutions to Einstein’s field equations. The creation and maintenance of traversable wormholes were first rigorously analysed by Morris and Thorne \cite{morris1988wormholes}, who explored the energy conditions that such structures must satisfy.
Their study revealed that traversable wormholes in general relativity (GR) inevitably violate the Weak Energy Condition (WEC), at least at the throat of the wormhole. Such a violation implies the presence of exotic matter —i.e. matter with negative energy density. Subsequent studies have confirmed that all static wormhole solutions in GR inherently require exotic matter for their support. This reliance on exotic matter is considered a major drawback, as it challenges the existence of wormholes within the framework of classical GR.\\
To analyse this, consider the energy-momentum tensor $T_{\mu\nu}$, which describes the distribution of energy and matter in spacetime. In the orthonormal frame basis, its diagonal components are defined as: $T^\mu_\nu=diag(-\rho(l),\tau(l),p(l),p(l)).$    
Assuming the Einstein field equations $G_{\mu\nu} = \kappa T_{\mu\nu}$ and $\kappa=1$ units, the energy-momentum tensor 
(written down from the metric and its Einstein tensor) of the generalised Bronnikov--Ellis wormhole with asymptotic mass is: 
\begin{align}
    \rho(l)= \rho_\Phi+\rho_e =&\left[-\frac{c_0^2 e^{f(l)}}{2\left(b_0^{\text{n}}+l^{\text{n}}\right){}^{\frac{4}{n}}}\right]
    +\Bigg[ e^{f(l)} \Bigg(\frac{c_0^2}{2\left(b_0^n+l^n\right){}^{\frac{4}{n}}}+\frac{2 l^{n-2} \left(l f'(l)-n+1\right)}{b_0^n+l^n}\nonumber\\
    &+ \frac{1}{\left(b_0^n+l^n\right){}^{\frac{2}{n}}}+\frac{ (2 n-3) l^{2 n-2}}{\left(b_0^n+l^n\right){}^2}+ f''(l)-\frac{f'(l)^2}{4}\Bigg)\Bigg] \label{emt1}
\end{align}
\begin{equation}
    \tau(l)=\tau_\Phi+\tau_e
     =\left[-\frac{c_0^2 e^{f(l)}}{2\left(b_0^{\text{n}}+l^{\text{n}}\right){}^{\frac{4}{n}}}\right] +  \left[ e^{f(l)}  \left(\frac{c_0^2}{2\left(b_0^n+l^n\right){}^{\frac{4}{n}}}-\frac{1}{\left(b_0^n+l^n\right){}^{\frac{2}{n}}}+\frac{l^{2 n-2}}{\left(b_0^n+l^n\right){}^2}-\frac{f'(l)^2}{4}\right)\right] \label{emt2}
\end{equation}
\begin{equation}
    p(l) = p_\Phi+p_e
     =\left[\frac{c_0^2 e^{f(l)}}{2\left(b_0^{\text{n}}+l^{\text{n}}\right){}^{\frac{4}{n}}}\right] +  \left[ e^{f(l)} \left(-\frac{c_0^2}{2\left(b_0^n+l^n\right){}^{\frac{4}{n}}}+\frac{ (n-1) b_0^n l^{n-2}}{\left(b_0^n+l^n\right){}^2}+\frac{f'(l)^2}{4}\right) \right]\label{emt3}
\end{equation}

\noindent It is clear that the energy densities and pressures change
with change in $n$ or $M$ ($C$). In each of the above expressions, the first term of the right-hand side is due to the phantom field, and the remaining terms denote the effects of extra matter. For the phantom field, the components of the energy-momentum tensor have the form
\begin{equation}
    \rho_\Phi=\tau_\Phi=-p_\Phi=-\frac{e^{f(l)}}{2}\Phi'^2 \hspace{1cm} \Phi'=\frac{d\Phi}{dl}
\end{equation}

\noindent Here the scalar field $\Phi(l)$ is found by solving the Klein-Gordon equation $\Box\Phi=0$. Interestingly, the $f(l)$ in the line element
does not appear in the KG equation. Solving for $\Phi(l)$, we get,
\begin{equation*}
    \Phi(l)=c_0\frac{l \,\, _2F_1[1/n,2/n,1+1/n,-(l/b_0)^n]}{b_0^2}
\end{equation*}
where ${}_2F_1[...]$ denotes the hypergeometric function. For $n=2$, $\Phi(l)=
\frac{c_0}{b_0}\arctan(\frac{l}{b_0})$--same as $f(l)$, after adjusting and adding
constants.
\noindent If $2c_0^2=C^2+4b_0^2$, then for the $n=2$ case, the contribution due to the extra term vanishes and the matter stress-energy is entirely due to the phantom scalar field.\\
Hence, the total action for this family of wormhole geometries
can be expressed as a sum of the Einstein-Hilbert gravity action ($S_g$), the phantom scalar action ($S_\Phi$) and an extra action ($S_{extra}$), the exact form of which is not known.
\begin{equation}
    S=S_g+S_\Phi+S_{extra}=\int \sqrt{-g}Rd^4x+\int \sqrt{-g}\partial^\nu\Phi\partial_\nu\Phi d^4x+S_{extra}
\end{equation}
\noindent Now, if we check the WEC conditions ($\rho\geq0$,\hspace{0.1cm}$\rho+\tau\geq0$ and $\rho+p\geq0$) in the $M=0$ case, we find that for $n=2$, all three conditions are not simultaneously satisfied in any region of $l$. For $n>2$, though the first and third conditions are satisfied in a specific range of $l$, the second one is always violated (except at $l=0$).
\begin{figure}[H]
    \centering
    {{\includegraphics[width=5.2cm]{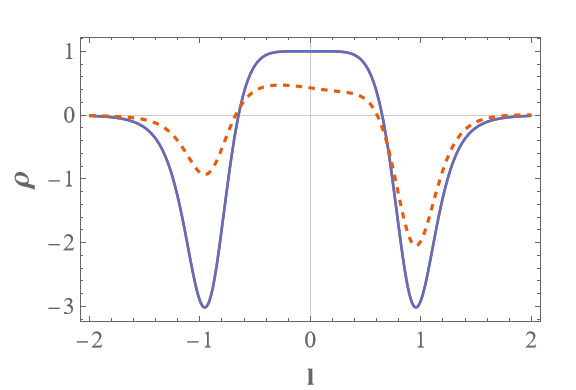} }}%
    \hspace{0.0cm}
    {{\includegraphics[width=5.2cm]{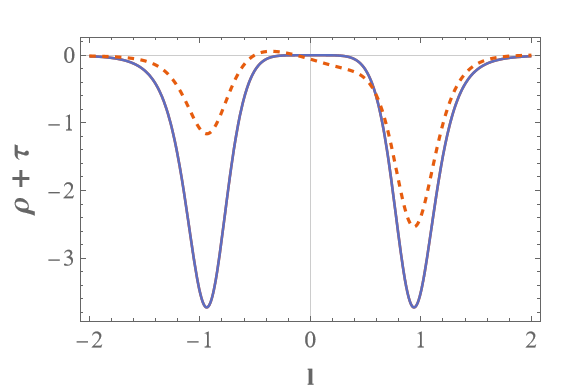} }}%
    \hspace{0.0cm}
    {{\includegraphics[width=5.2cm]{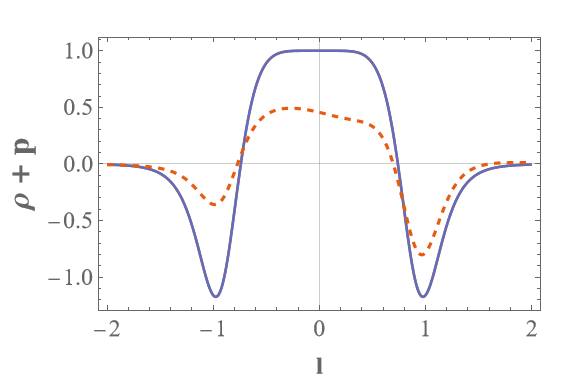} }}%
    \caption{\footnotesize For $n=8$ geometry: [left] $\rho$; [middle] $\rho+\tau$; [right] $\rho+p$. Blue solid line denotes $M=0$ and orange dashed line represents $M=0.25b_0$ (here $b_0=1$). In all the images, the x-axis represents the proper coordinate $l$.}
    \label{fig:StressEnergy4_123}
\end{figure}

\noindent However, in the case of $M\neq0$ and $n>2$, one can find that all three conditions are satisfied in certain regions if the $|M|$ exceeds a specific value (Fig. \ref{fig:StressEnergy4_123}). For different geometries (different values of $n$), the critical value of $M$ and the acceptable regions (corresponding to a fixed value of $M$) are different.\\

\subsection{Bronnikov--Ellis geometry with a chosen redshift function:} \label{chosen_redshift_func}

\noindent 
It is also possible to choose a suitable redshift function ($-g_{00}$ never zero) in an ad-hoc way while retaining the Bronnikov--Ellis $n=2$ character unaltered in the spacelike slice. The ensuing consequences
for required matter and energy conditions can thereafter be looked at. 
As an example, let us consider the line element studied in \cite{alfaro2024quasinormal} and given by,
\begin{equation*}
    e^{2\Psi(r)}=1-\frac{2M}{r} \hspace{2cm} b(r)=\frac{b_0^2}{r}
\end{equation*}
where $b_0> 2M$ is required to make sure that there is no horizon and we have a wormhole.
\noindent Thus, we have the modified  Bronnikov--Ellis geometry with line element given as,
\begin{equation}
    ds^2 = - \left(1-\frac{2M}{r}\right)dt^2 + \frac{dr^2}{1-\frac{b_0^2}{r^2}} + r^2 d\theta^2 + r^2 sin^2\theta d\varphi^2
\end{equation}
For this modified geometry, the components of the energy-momentum tensors ($T^\mu_\nu=diag(-\rho(r),\tau(r),p(r),p(r))$) have the form \cite{alfaro2024quasinormal}
\begin{equation}
    \rho(r)=-\frac{b_0^2}{r^4}, \hspace{1.5cm} \tau(r)=\frac{2Mr-b_0^2}{r^3(r-2M)}, \hspace{1.5cm} p(r)=\frac{(M-r) \left(b_0^2 (M-r)+M r^2\right)}{r^4 (r-2 M)^2}
\end{equation}

\begin{figure}[H]
    \centering
    {{\includegraphics[width=5.2cm]{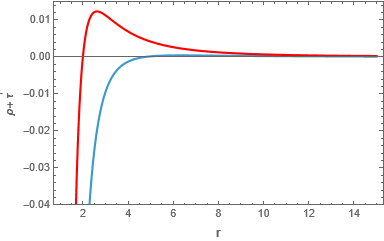} }}%
    \hspace{1.0cm}
    {{\includegraphics[width=5.2cm]{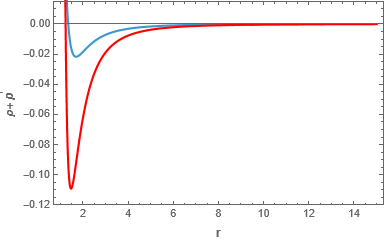} }}%
    \caption{\footnotesize For Bronnikov--Ellis wormhole with redshift function: [left] $\rho+\tau$; [right] $\rho+p$. Blue line denotes $M=0.2b_0$ and red line represents $M=0.4b_0$. (here $b_0=1$)}%
    \label{fig:StressEnergy_23}%
\end{figure}

\noindent Note that the energy density is not affected by $M$, though the pressures change.
The Weak Energy Condition is violated for all $r>b_0$ since the energy density is negative. The Null Energy Condition is also violated
with complementary behaviour for the $\rho+\tau$ and $\rho+p$ inequalities w.r.t. increase in $M$ value (Fig. \ref{fig:StressEnergy_23}). As $M$ increases the 
region of violation (near $r=b_0$) of the $\rho+\tau$ inequality  shrinks whereas the region of violation (away from $r=b_0$) of the $\rho+p$
inequality becomes larger. We have plotted $\rho+\tau$ and $\rho+p$ as a function of the radial coordinate $r$, for two values of $M$ (Fig. \ref{fig:StressEnergy_23}), in order to show this feature in the
NEC violations.

\noindent Putting $M=0$, we get back the original Bronnikov--Ellis wormhole. We will discuss the effective potential and the $\omega_{QNM}$ for this spacetime in a later section (\ref{Mod_Ellis}).

\subsection{Effective potentials for scalar and axial perturbation} \label{eff}

\noindent When an open system is perturbed, it emits radiation and eventually stabilises. QNMs represent the damped oscillations of perturbations in spacetime and can arise due to scalar, electromagnetic or gravitational perturbations.\\
The original Bronnikov--Ellis wormhole ($n=2$) is a solution to the Einstein equations coupled to a phantom scalar field. The $n>2$ geometries also have contributions from the phantom field. The perturbations in this phantom scalar field ($\Phi$) can give rise to scalar quasinormal modes. For this, we may write the scalar field as $\Phi+\phi$. Here $\phi$ is the perturbation on the phantom scalar field 
and it satisfies
\begin{equation}
    \Box \Phi= \frac{1}{\sqrt{-g}}\partial_\mu(\sqrt{-g}g^{\mu\nu}\partial_{\nu}\Phi)=0 \label{kg}
\end{equation}
Putting the ansatz $\Phi= \frac{e^{f(l)/2}}{(l^n+b_0^n)^{1/n}}\psi(l)\hspace{0.1cm}e^{-i\omega t}Y(\theta,\varphi)$ (here $Y(\theta,\varphi)$ are the spherical harmonics; stationary ansatz: $e^{-i\omega t}$ ) in Eqn. \eqref{kg}, one gets the Schrodinger-like radial equation 
\begin{equation}
    \frac{d^2 \psi}{d l_*^2} +(\omega^2- V_{sc}(l)) \psi = 0 \label{master} \hspace{2cm} \text{where} \hspace{1cm} \frac{dl_*}{dl}=e^{-f}
\end{equation}
Here, $V_{sc}$ is the effective potential (scalar perturbation) for the massive generalised Bronnikov--Ellis wormhole (Eq. \eqref{AsymM_Ellis}), given by
\begin{equation}
    V_{sc}(l)=\frac{1}{4} e^{2 f(l)} \left(\frac{4 m \left(m+1\right)}{\left(l^n+b_0^n\right){}^{2/n}}+\frac{4 (\text{n}-1) b_0^{\text{n}} l^{\text{n}-2}}{\left(l^n+b_0^n\right){}^2}-2 f''(l)-f'(l)^2\right) \label{vsc}
\end{equation}
where $m$ is the azimuthal number (from now on, will be referred to as the angular momentum parameter).\\
In linear perturbation theory, gravitational perturbation is classified into two categories based on parity: axial (odd parity) and polar (even parity). In order to get the master equation for axial perturbation, we follow Chandrasekhar's method \cite{Chandrasekhar1998}. Finally, we get
\begin{equation}
    \frac{d^2 \psi}{d l_*^2} +(\omega^2- V_{ax}(l)) \psi = 0 \label{master_ax}
\end{equation}
where the effective potential has the form:

\begin{equation}
    V_{ax}=\frac{1}{4} e^{2 f(l)} \left(-\frac{4 l^{n-2} \left(2 l f'(l)+n-1\right)}{l^n+b_0^n}+\frac{4 \left(m^2+m-2\right)}{\left(l^n+b_0^n\right){}^{2/n}}+\frac{4 (n+1) l^{2 n-2}}{\left(l^n+b_0^n\right){}^2}+2 f''(l)+3 f'(l)^2\right) \label{vax}
\end{equation}
Plotting the potentials in the ultrastatic case ($M=0$) reveals that both $V_{sc}$ and $V_{ax}$ have a single-peaked nature (Fig. \ref{N2_pot_sc},\ref{fig:N2_pot}). One can notice that if $n>2$, $V_{sc}$ has double barriers for all values of the angular momentum parameter (m). The peaks move towards the origin and become less prominent with increasing m (Fig. \ref{fig:N4_pot_sc}). The effective potential due to axial perturbation exhibits a triple barrier up to a certain value of m (this critical value of m is different for different n values). If we further increase m, we will first observe potentials with steps and then single barriers with flattened peaks (See Fig. \ref{fig:N8_pot}).

\noindent In the $M\neq0$ cases, for $n=2$ geometry, both axial and scalar potentials have asymmetric single peaks that get shifted from the origin. The shift and asymmetry are also noticeable for $n>2$ cases, where the height of the peaks on the positive side of $l_*$ exceeds that of the negative side (if $M>0$).

\section{Results: quasinormal spectra (fundamental and overtones for the ultra-static case)} \label{result}

\noindent The problem of computing quasi-normal modes for the Bronnikov--Ellis wormhole has been tackled in many previous works. The authors of \cite{kim2008wormhole,blazquez2018scalar} and \cite{dutta2020revisiting,dutta2022novel} have presented the fundamental modes for the original Bronnikov--Ellis ($n=2$ case) and the generalised Bronnikov--Ellis wormholes, respectively.

\noindent However, as pointed out in \cite{Sago2021,berti2007matched,giesler2019black,nee2023role}, the fundamental modes may not represent the full picture, as the overtones play a crucial role in fitting the ringdown (RD) waveform. The inclusion of overtones (at least, the first few overtones) can help us achieve a better fit of the RD profile at earlier times when we expect a better signal-to-noise ratio (SNR). Overtones can also be instrumental in providing better estimates of the parameters of the underlying geometry. Though the overtones corresponding to the $n=2$ geometry were computed in \cite{alfaro2024quasinormal,batic2024unified}, we compute the full QNM spectra for the generalised Bronnikov--Ellis spacetimes (even $n>2$ cases). We note many key differences between the $n=2$ and the $n>2$ cases.

\noindent The authors of \cite{baibhav2023agnostic} have argued that the overtones having lower real parts compared to the fundamental mode (true for $n=2$), fit away the non-linearities of the RD signal. The same argument will not be true for $n>2$ geometries (where Re$(\omega)$ for overtones are higher in value than the fundamental mode). Any possible future detection of such RD signals may further cement the role of overtones.

\noindent While computing the $\omega_{QNM}$, we have used a frequency-domain method (spectral method (SM)) in conjunction with a time-domain method (Prony fit) (see appendix \ref{method}). While the frequency-domain method is indispensable for extracting all the modes (up to a certain overtone number), the time-domain provides us with some valuable insight into the relative amplitudes of the overtones, thus pointing to the fact that sometimes even higher-order overtones (in our case: even modes) can have greater importance than the lower-order (in our case: odd modes) ones.
 
\noindent The results from the two methods are in good agreement with each other. However, we could not compute the same number of overtones for every geometry. For lower values of $n$ and angular momentum parameter $m$, we could get very few overtones. 

\noindent In the series of figures below, we demonstrate the effective potentials and QNMs for scalar and axial gravitational perturbations.
Figs. 3-6 are for $n=2$, Figs. 7-10 for $n=4$, Figs. 11-14 are for 
$n=6$ and Figs. 15-18 for $n=8$. In the following section, we discuss the results here and
try to arrive at specific conclusions on the nature of the QNMs and
their link with the different geometries. 

\subsection*{QNMs for n = 2 geometry:}
\begin{figure}[H]
\centering
\parbox{7.5cm}{
\includegraphics[width=7.5cm]{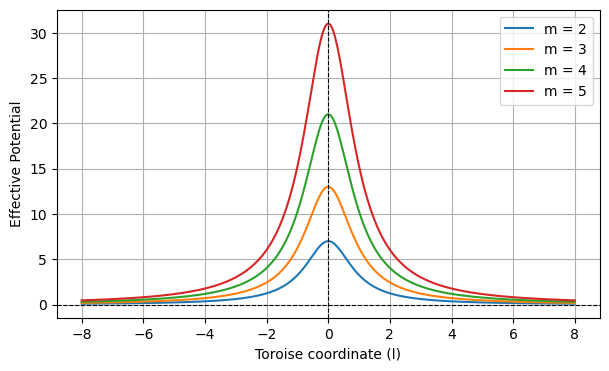}
\caption{\footnotesize Effective potential (Scalar) for Bronnikov--Ellis wormhole with $n=2$.}
\label{N2_pot_sc}}
\qquad
\begin{minipage}{7.5cm}
\includegraphics[width=7.5cm]{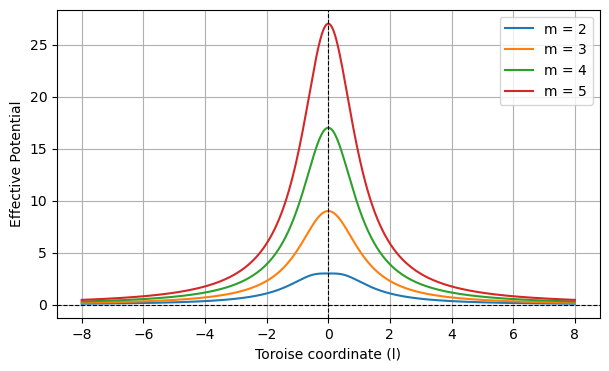}
\caption{\footnotesize Effective potential (Axial) for Bronnikov--Ellis wormhole with $n=2$.}
\label{fig:N2_pot}
\end{minipage}
\end{figure}

\begin{figure}[H]
\centering
\parbox{7.5cm}{
\includegraphics[width=7.5cm]{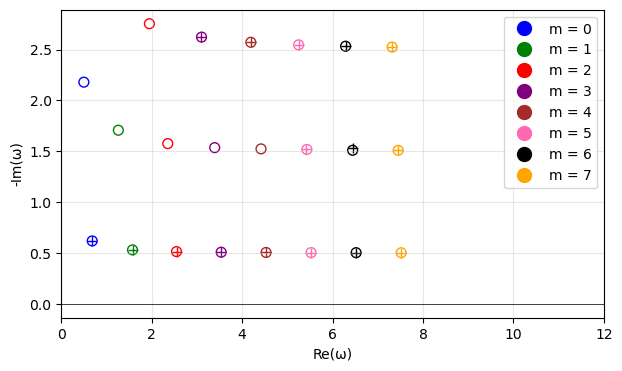}
\caption{\footnotesize QNMs for n = 2 (Scalar) case. The results of SM and Prony fit are represented by circles and plus signs, respectively. (Table: \ref{tab:qnm_scalar_axial_2})}
\label{fig:N2_sc}}
\qquad
\begin{minipage}{7.5cm}
\includegraphics[width=7.5cm]{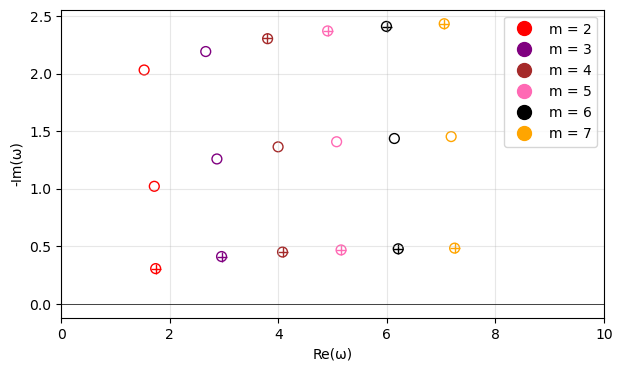}
\caption{\footnotesize QNMs for n = 2 (Axial) case. The results of SM and Prony fit are represented by circles and plus signs, respectively. (Table: \ref{tab:qnm_scalar_axial_2})}
\label{fig:N2_ax}
\end{minipage}
\end{figure}

\subsection*{QNMs for n = 4 geometry:}

\begin{figure}[H]
\centering
\parbox{7.5cm}{
\includegraphics[width=7.5cm]{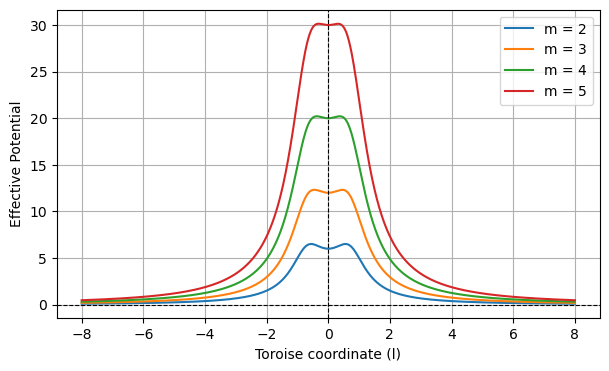}
\caption{\footnotesize Effective potential (Scalar) for generalised Bronnikov--Ellis wormhole with $n=4$.}
\label{fig:N4_pot_sc}}
\qquad
\begin{minipage}{7.5cm}
\includegraphics[width=7.5cm]{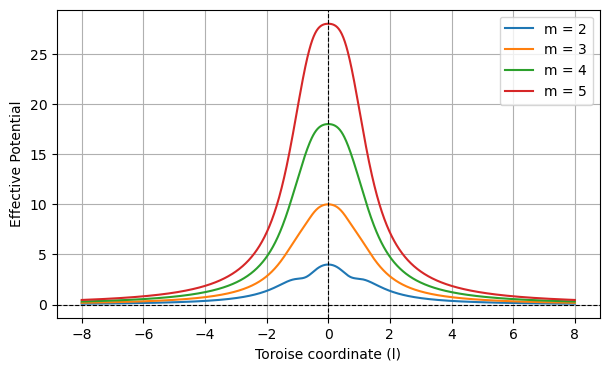}
\caption{\footnotesize Effective potential (Axial) for generalised Bronnikov--Ellis wormhole with $n=4$.}
\label{fig:N4_pot}
\end{minipage}
\end{figure}

\begin{figure}[H]
\centering
\parbox{7.5cm}{
\includegraphics[width=7.5cm]{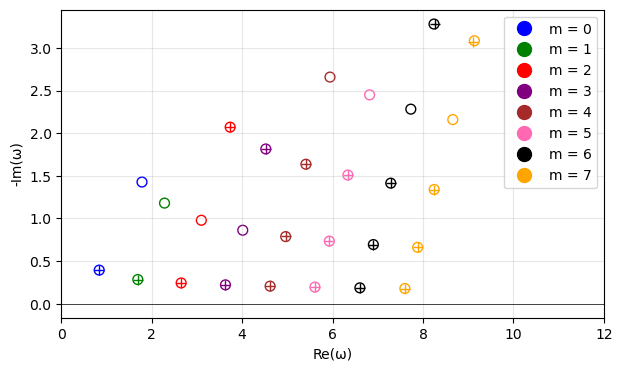}
\caption{\footnotesize QNMs for n = 4 (Scalar) case. The results of SM and Prony fit are represented by circles and plus signs, respectively. (Table: \ref{tab:qnm_scalar_axial_4})}
\label{fig:N4_sc}}
\qquad
\begin{minipage}{7.5cm}
\includegraphics[width=7.5cm]{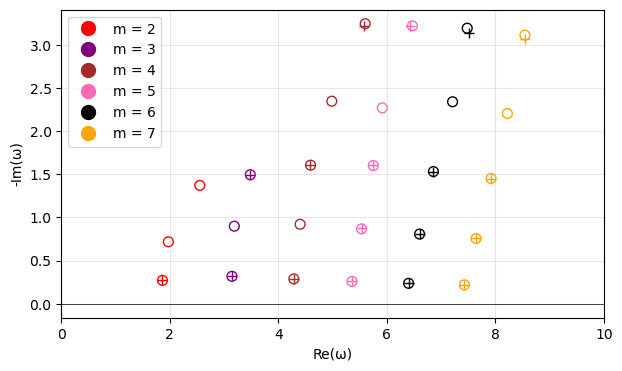}
\caption{\footnotesize QNMs for n = 4 (Axial) case. The results of SM and Prony fit are represented by circles and plus signs, respectively. (Table: \ref{tab:qnm_scalar_axial_4})}
\label{fig:N4}
\end{minipage}
\end{figure}

\subsection*{QNMs for n = 6 geometry:}

\begin{figure}[H]
\centering
\parbox{7.5cm}{
\includegraphics[width=7.5cm]{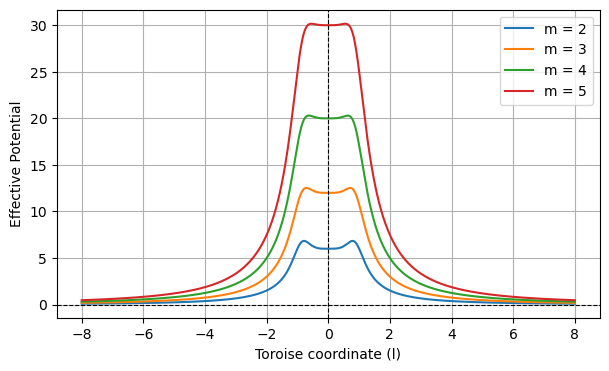}
\caption{\footnotesize Effective potential (Scalar) for generalised Bronnikov--Ellis wormhole with $n=6$.}
\label{fig:N6_pot_sc}}
\qquad
\begin{minipage}{7.5cm}
\includegraphics[width=7.5cm]{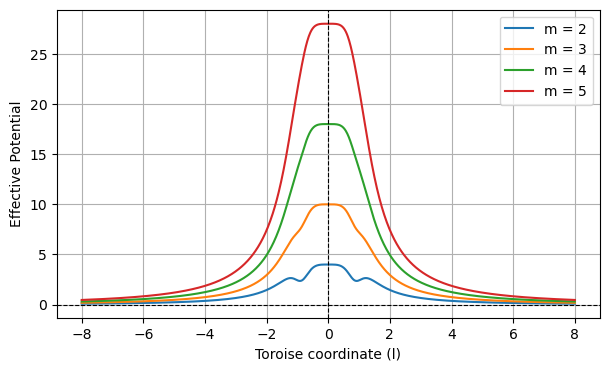}
\caption{\footnotesize Effective potential (Axial) for generalised Bronnikov--Ellis wormhole with $n=6$.}
\label{fig:N6_pot}
\end{minipage}
\end{figure}

\begin{figure}[H]
\centering
\parbox{7.5cm}{
\includegraphics[width=7.5cm]{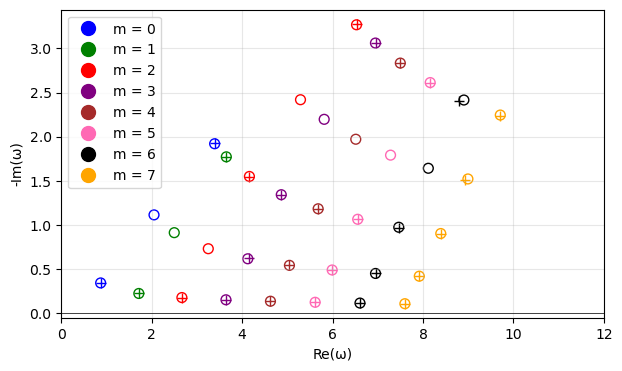}
\caption{\footnotesize QNMs for n = 6 (Scalar) case. The results of SM and Prony fit are represented by circles and plus signs, respectively. (Table: \ref{tab:qnm_scalar_axial_6})}
\label{fig:N6_sc}}
\qquad
\begin{minipage}{7.5cm}
\includegraphics[width=7.5cm]{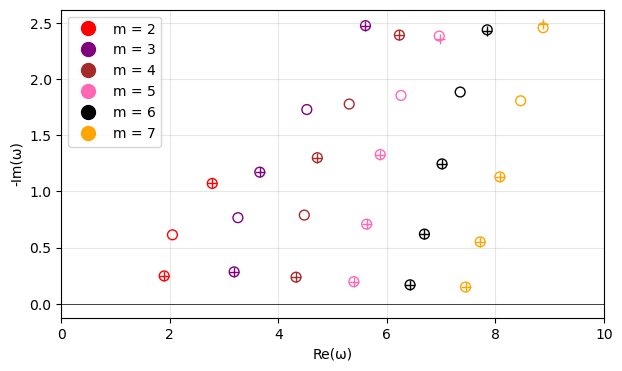}
\caption{\footnotesize QNMs for n = 6 (Axial) case. The results of SM and Prony fit are represented by circles and plus signs, respectively. (Table: \ref{tab:qnm_scalar_axial_6})}
\label{fig:N6}
\end{minipage}
\end{figure}

\subsection*{QNMs for n = 8 geometry:}

\begin{figure}[H]
\centering
\parbox{7.5cm}{
\includegraphics[width=7.5cm]{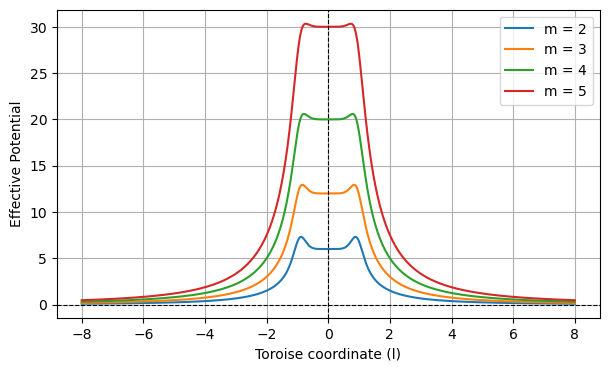}
\caption{\footnotesize Effective potential (Scalar) for generalised Bronnikov--Ellis wormhole with $n=8$.}
\label{fig:N8_pot_sc}}
\qquad
\begin{minipage}{7.5cm}
\includegraphics[width=7.5cm]{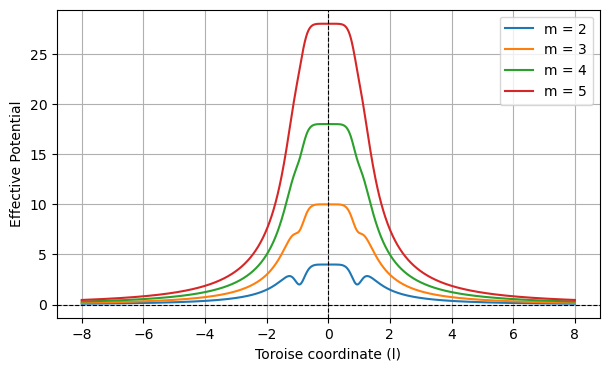}
\caption{\footnotesize Effective potential (Axial) for generalised Bronnikov--Ellis wormhole with $n=8$.}
\label{fig:N8_pot}
\end{minipage}
\end{figure}

\begin{figure}[H]
\centering
\parbox{7.5cm}{
\includegraphics[width=7.5cm]{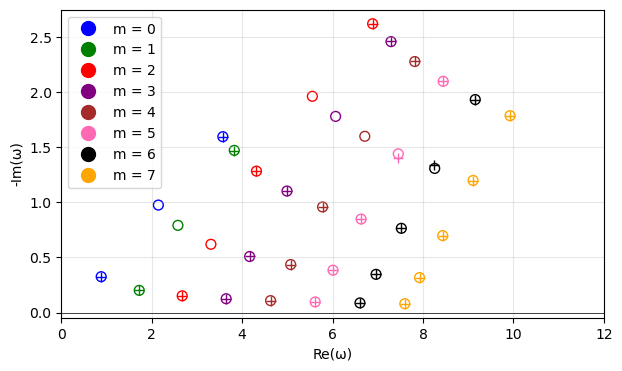}
\caption{\footnotesize QNMs for n = 8 (Scalar) case. The results of SM and Prony fit are represented by circles and plus signs, respectively. (Table: \ref{tab:qnm_scalar_axial_8})}
\label{fig:N8_sc}}
\qquad
\begin{minipage}{7.5cm}
\includegraphics[width=7.5cm]{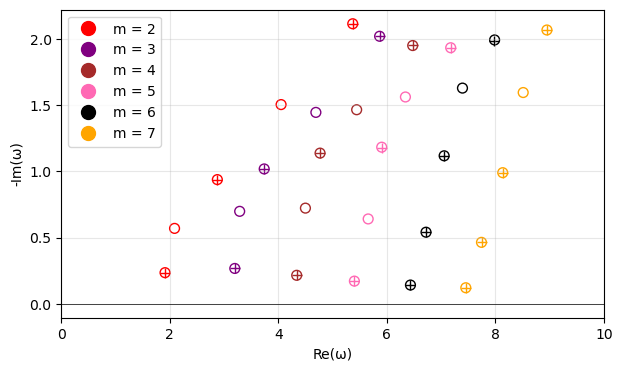}
\caption{\footnotesize QNMs for n = 8 (Axial) case. The results of SM and Prony fit are represented by circles and plus signs, respectively. (Table: \ref{tab:qnm_scalar_axial_8})}
\label{fig:N8}
\end{minipage}
\end{figure}

\section{Inferences drawn from the QNM spectra: emphasis on overtones} \label{Inference}

\noindent In Section \ref{result}, we presented the fundamental modes and overtones of the generalised, ultrastatic, Bronnikov--Ellis wormhole. Now, one may ask whether these QNMs can help distinguish different geometries (i.e. different $n$ values) of this wormhole family. To answer this query, we focus on the plots (Re$(\omega)$ vs $-Im(\omega)$) given above
in the previous section and try to identify some general trends in the QNM spectra. In our analysis, we denote the mode numbers with $p$, where the mode with the smallest value of $|Im(\omega)|$ (the longest-lived mode) is the fundamental mode (see e.g. \cite{giesler2019black}) with $p=0$ and modes with higher $|Im(\omega)|$ are the overtones.

\noindent For both scalar and axial perturbations, in the $n=2$ case, the real parts of the $\omega_{QNM}$ decrease with increase in mode number ($p$), while for the geometries with higher $n$ values, the higher overtones correspond to an increase in Re$(\omega)$.

\noindent From the tables in Appendix \ref{Table}, we notice that for a fixed large value of the angular momentum parameter ($m$), the fundamental modes of higher $n$ geometries are very close in value. However, these wormhole spacetimes can also be distinguished from each other if we take the overtones into account. As we move to larger $n(>2)$ values, we observe that the rate of increase of Re$(\omega)$ with the mode number gets expedited. This is true for both scalar and axial perturbations (see Figs. \ref{fig:N2_sc}, \ref{fig:N2_ax}, \ref{fig:N4_sc}, \ref{fig:N4}..., \ref{fig:N8}).

\noindent The imaginary parts of the quasinormal spectra are also an important feature distinguishing the geometries. In the scalar perturbation case, for a fixed angular momentum parameter value, the greater the $n$ value, the smaller the magnitude of the imaginary parts of the QNMs. The separation between the $|Im(\omega)|$ also keeps decreasing (See Figs. \ref{fig:N2_sc}, \ref{fig:N4_sc}, \ref{fig:N6_sc}, \ref{fig:N8_sc}). This implies that for wormholes with larger values of $n$, the fundamental modes and overtones have a lower decay rate and, therefore, are less stable. As we keep on increasing $m$, for the axial perturbation case, the absolute value of the imaginary parts of the fundamental modes first increase and then decrease (See Figs. \ref{fig:N2_ax}, \ref{fig:N4}, \ref{fig:N6}, \ref{fig:N8}).

\begin{table}[H]
    \centering
    \setlength{\tabcolsep}{10pt}  
    \renewcommand{\arraystretch}{1.2}  
    \singlespacing
    \begin{tabular}{c c c}
        \hline
        \hline
        Variation & Scalar & Axial \\
        \hline
         &The rate of increase &The rate of increase \\
        \textbf{Increase n:} &  of Re$(\omega)$ with the mode   & of Re$(\omega)$ with the mode \\ 
        (for $n>2$) & number gets expedited. & number gets expedited. \\
        \hline
         &Re$(\omega)$ increases and &Re$(\omega)$ increases and\\
        \textbf{Increase m:} &  $|Im(\omega)|$ decreases  & $|Im(\omega)|$ (fundamental mode) first  \\ 
         & &increases and then decreases. \\
         \hline
         &The rate of increase of &The rate of increase of\\
        \textbf{Increase p:} &   Re$(\omega)$ does not change &  Re$(\omega)$ increases after a\\ 
        (for $n>2$) &   with overtone number. & certain overtone number. \\
         &   $|Im(\omega)|$ increases. & $|Im(\omega)|$ increases. \\
        \hline
        \hline
    \end{tabular}
    \caption{\footnotesize Key features of the QNM spectra for massless generalised Bronnikov-Ellis wormhole.}
    \label{tab:tab_key}
\end{table}

\noindent Having identified the broader trends from the plots, we can try to explain them from the effective potential. The decrease of Re$(\omega)$ with overtone mode number is only observed in the $n=2$ case, which has a comparatively narrower and sharper single-peaked potential (See Figs. \ref{N2_pot_sc}, \ref{fig:N2_pot}). This trend is reversed for all even $n>2$ wormholes, which have a comparatively broader and flatter effective potential with two peaks for scalar perturbation (See Figs. \ref{fig:N4_pot_sc}, \ref{fig:N6_pot_sc}, \ref{fig:N8_pot_sc}) and one (for higher values of m)/three (for smaller values of m) peak(s) for axial perturbations (See Figs. \ref{fig:N4_pot}, \ref{fig:N6_pot}, \ref{fig:N8_pot}). On careful observation, one can notice that as we go to higher values of $n$, the central region of the potential becomes flatter. This can be a possible explanation behind the enhanced rate of increase of the real parts for higher $n$ geometries.

\noindent Fixing the value of $n$, if we increase the angular momentum parameter ($m$), the height of the effective potential increases, which is related to increasing the real part of the QNMs. The double peak feature also becomes less prominent for the scalar perturbation case (See Figs. \ref{fig:N4_pot_sc}, \ref{fig:N6_pot_sc}, \ref{fig:N8_pot_sc}). This may explain the decrease of $|Im(\omega)|$ with $m$ in the quasi-normal spectra. For axial perturbations in $n>2$ geometries, the transition of the potential from triple barrier to step-like and then to a single barrier, with increasing m (See Figs. \ref{fig:N4_pot}, \ref{fig:N6_pot}, \ref{fig:N8_pot}), may cause the initial increase and subsequent decrease of $|Im(\omega)|$ for the fundamental modes with m.

\subsection{Approximate Analytic Fit}

\noindent To further advance our study of the QNM spectra and their dependence on different parameters such as the geometry ($n$), angular momentum parameter ($m$), overtone order ($p$) and the throat ($b_0$), we can fit our numerical results (for scalar perturbations) to analytical models. \\
Here, the throat radius is simply a scaling parameter, and the quasi-normal frequencies are inversely proportional to $b_0$. However, the other parameters do not follow such simple rules. To get the fits corresponding to these parameters, we have used the NonLinearModelFit function of {\em Mathematica 14.2}. For all the fits obtained in this section, we have used the results from the spectral method.\\
As mentioned above, the change in the QNMs with changing $n$ values may give us some valuable insights. To this end,
we have analysed the dependence of Re$(\omega)$ on overtone mode number ($p$) and the angular momentum parameter ($m$), keeping the geometry parameter $n$ fixed. 
\begin{figure}[H]
    \centering
    \includegraphics[width=7.0cm]{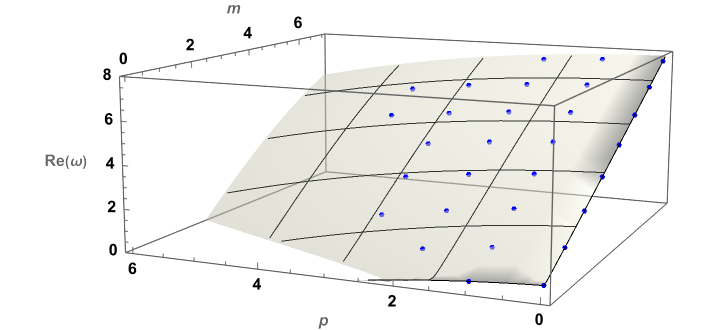}
    \caption{\footnotesize QNMs for n = 2 (Scalar Perturbation) case.}
    \label{fig: Nn2}
\end{figure}

\noindent Here, we observe that the real parts of the quasi-normal frequencies decrease with overtone mode $p$ but increase with angular momentum parameter $m$ (Fig. \ref{fig: Nn2}).

\begin{figure}[H]
    \centering
    {{\includegraphics[width=7.0cm]{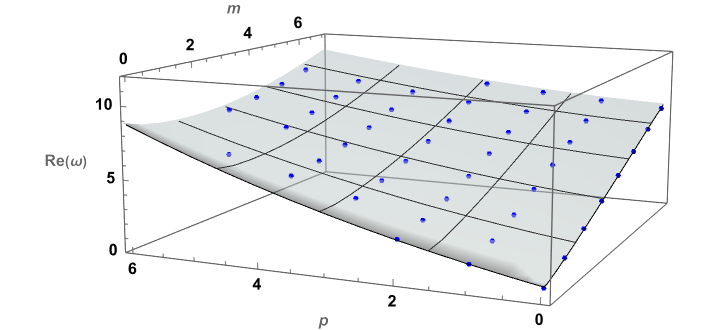} }}%
    {{\includegraphics[width=7.0cm]{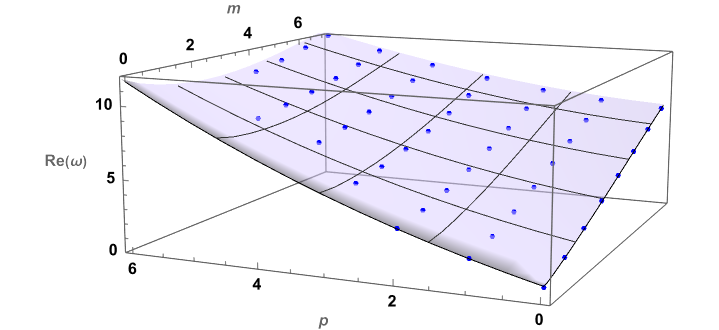} }}%
    \caption{\footnotesize [Left] QNMs for n = 4; [Right] QNMs for n = 8.}%
    \label{fig:fit_4}%
\end{figure}

\begin{figure}[H]
    \centering
    {{\includegraphics[width=5.2cm]{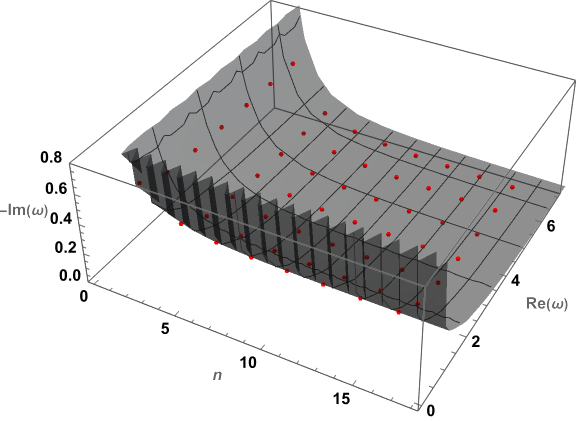} }}%
    {{\includegraphics[width=5.2cm]{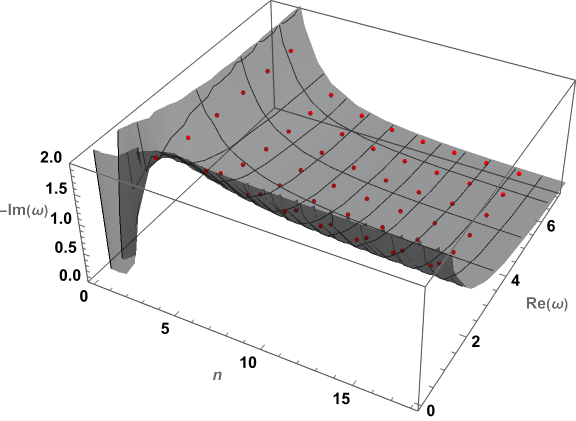} }}%
    {{\includegraphics[width=5.2cm]{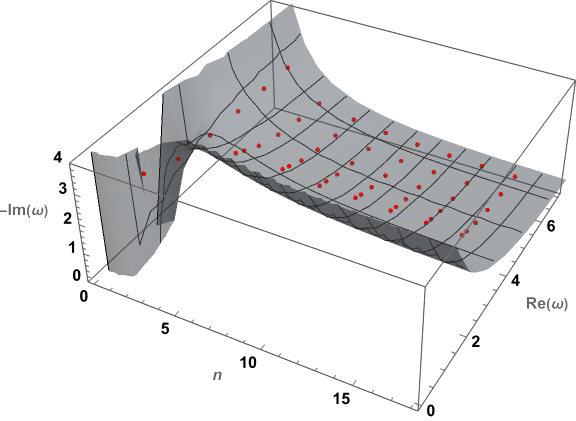} }}%
    \caption{\footnotesize Fit of Im($\omega$) with n and Re($\omega$): [left] fundamental modes; [middle] 1st overtones; [right] 2nd overtones.}%
    \label{fig:fit_123}%
\end{figure}

\noindent For geometries with higher $n$ $(>2)$, the real parts increase with $m$ as well as with overtone mode number $(p)$. We notice, with increasing $p$ Re$(\omega)$ increases faster for $n=8$ case as compared to say, the $n=4$ case (Fig. \ref{fig:fit_4}).
We have also fitted the imaginary part with the geometry 
parameter and the real part, for different overtone numbers ($p$), for scalar perturbations (Fig. \ref{fig:fit_123}).

\noindent Thus far, we have (i) fitted Re$(\omega_{QNM})$ with angular momentum and overtone mode number and  (ii) the $- Im(\omega_{QNM})$ with geometry parameter $n$ and Re$(\omega_{QNM})$. Hence, for given values of $n$, $m$ and $p$, we can numerically estimate the value of $\omega_{QNM}$ (scalar perturbation). The fitting functions discussed so far are approximate. They do not return exact results but provide information about overall trends, which are 
useful while interpreting our results.

\subsection{Detectable Modes}

\noindent Since we have now obtained the quasinormal modes including some overtones, it is important to ask whether or not these modes are detectable. If they are within observable limits, this must be for 
certain choices of the parameters involved. To figure this out, we will need to determine the numerical values for the quasinormal frequencies ($\nu$) and decay times ($\tau$), which can be found from
\begin{equation*}
    \nu = \frac{c^3}{2 \pi b_0 G M_\odot}Re(\omega) \hspace{0.2cm} Hz  \hspace{2cm} \tau = \frac{2 \pi b_0 G M_\odot}{c^3}\frac{1}{Im(\omega)} \hspace{0.2cm} seconds
\end{equation*}

\noindent Advanced LIGO is sensitive to the frequency range of $10 \hspace{0.1cm} Hz$ to $10 \hspace{0.1cm} kHz$. For the generalised 
Bronnikov-Ellis geometries, we have calculated the QNFs, and all 
of them can be within the aLIGO frequency range when the throat radius is roughly between $50M_\odot$ and $1000M_\odot$. For now, we choose $b_0=200$ (in units of solar mass) and present the frequencies and damping times for a few cases, as an illustration. In Table \ref{tab:tab25} below, we show the values for $n=2$ and $m=5$ (scalar perturbation).
\begin{table}[H]
    \centering
    \singlespacing
    \begin{tabular}{|c|c|c|c|}
        \hline
        Mode & Frequency (Hz) & Damping Time (s) & Amplitude\\
        \hline
        0 & 889.5 & 0.01235 &  0.2396 \\ 
        \hline
        1 & 872.6 & 0.00408 & 2.01$\times 10^{-5}$ \\ 
        \hline
        2 & 845.8 & 0.00244 & 0.1219 \\ 
        \hline
    \end{tabular}
    \caption{\footnotesize Frequency and Damping time for $n=2$ and $m=5$. These values are calculated from the Prony fit (details in Appendix \ref{Table}).}
    \label{tab:tab25}
\end{table}
\begin{figure}[H]
\centering
\parbox{7.5cm}{
\includegraphics[width=7.3cm]{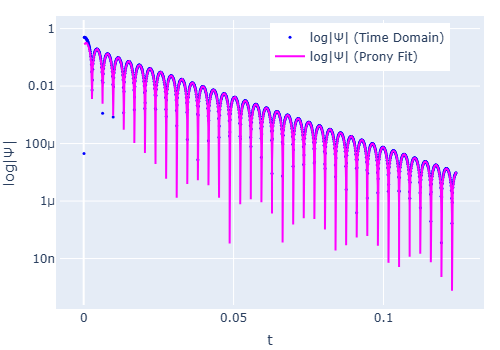}
\caption{\footnotesize Log plot of time-domain profile for $\psi$ for $n=2$ and $m=5$.}
\label{ln1}}
\qquad
\begin{minipage}{7.5cm}
\includegraphics[width=7.3cm]{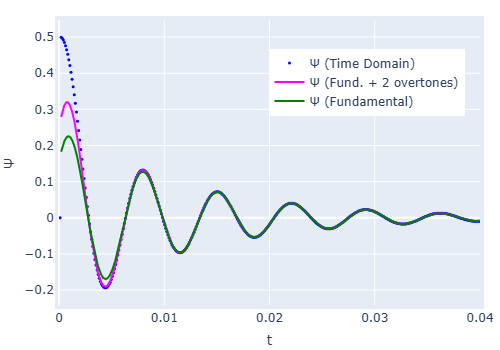}
\caption{\footnotesize Time-domain waveform for $n=2$ and $m=5$.}
\label{fig:2figsB}
\end{minipage}
\end{figure}

\noindent Frequencies and damping
times for n = 8 and m = 5 case (scalar perturbation) are mentioned in Table \ref{tab:tab85}
\begin{table}[H]
    \centering
    \singlespacing
    \begin{tabular}{|c|c|c|c|}
        \hline
        Mode & Frequency (Hz) & Damping Time (s) & Amplitude\\
        \hline
        0 & 904.2 & 0.06467 &  0.1550 \\ 
        \hline
        1 & 967.6 & 0.01617 & 1.789$\times 10^{-5}$ \\ 
        \hline
        2 & 1067.7 & 0.00733 & 0.1368 \\ 
        \hline
        3 & 1200.3 & 0.00444 & 0.0006 \\ 
        \hline
        4 & 1359.8 & 0.00296 & 0.1100 \\ 
        \hline
    \end{tabular}
    \caption{\footnotesize Frequency and Damping time for $n=8$ and $m=5$. These values are calculated from the prony fit (details mentioned in Appendix \ref{Table}).}
    \label{tab:tab85}
\end{table}

\begin{figure}[H]
\centering
\parbox{7.5cm}{
\includegraphics[width=7.5cm]{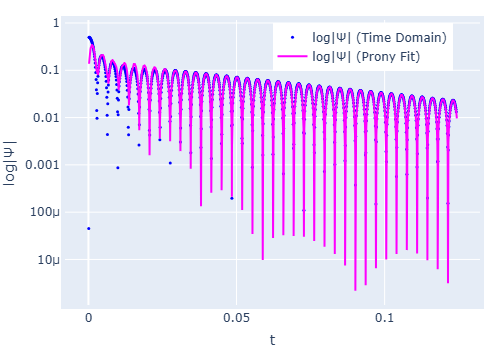}
\caption{\footnotesize Log plot time-domain profile for $\psi$ for $n=8$ and $m=5$.}
\label{ln}}
\qquad
\begin{minipage}{7.5cm}
\includegraphics[width=7.5cm]{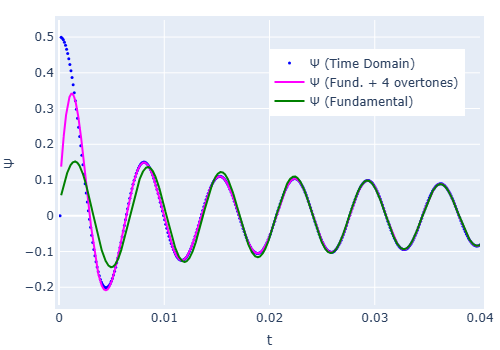}
\caption{\footnotesize Time-domain waveform for $n=8$ and $m=5$.}
\label{fig:8figsB}
\end{minipage}
\end{figure}

\noindent In the ringdown waveforms, we have represented the actual time-domain (obtained using the procedures mentioned in \ref{Prony}) wave as the blue points, the fundamental mode as green lines and the resultant wave of the fundamental and overtones ($\omega$ and amplitudes computed using the Prony fit) as the magenta lines (Figs. \ref{fig:2figsB},\ref{fig:8figsB}). For both cases, the magenta lines better fit the time domain than only the fundamental mode. However, the difference between the green and magenta lines is only noticeable for the first few cycles, as the overtones have higher imaginary parts and decay faster. While the inclusion of overtones makes little improvement in the fitting for the $n=2$ case, the effect is more significant for the $n=8$ case. This is because for the $n=8$ case, we could find more overtones, and these overtones have amplitudes comparable to the fundamental mode, whereas, in the $n=2$ case, the amplitudes of the overtones are significantly lower than the fundamental mode (here the fundamental mode fits well from the second cycle) . For $n=4,6$ cases, we computed $2-3$ overtones and their corresponding amplitudes using the Prony fit, and similar to the $n=8$ case, we observe that inclusion of the overtones significantly improves the fit at early times. For all the geometries (i.e. different $n$), the more overtones we could get (up to $p=4$), the better we were able to model the time-domain signal, indicating that obtaining more overtones may result in a better fit of the time-domain profile. However, we need to be cautious not to overfit our results by incorporating an excessive number of overtones.

\begin{figure}[H]
\centering
\parbox{6cm}{
\includegraphics[width=6cm]{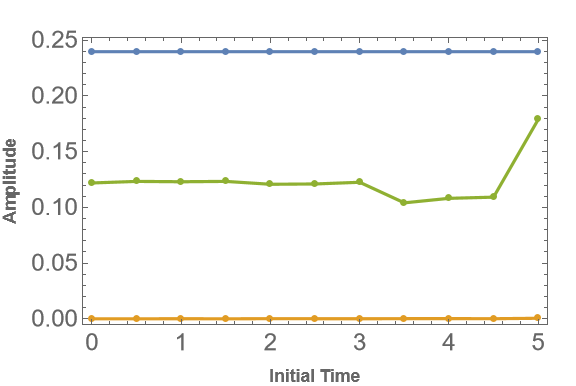}
\caption{\footnotesize Amplitude calculated for $n=2$ and $m=5$ with different initial time.}
\label{fig:3figsA}}
\qquad
\begin{minipage}{2cm}
\includegraphics[width=2cm]{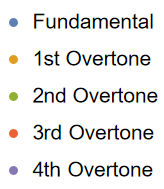}
\captionsetup[figure]{labelformat=empty}
\label{fig:3figsC}
\end{minipage}
\qquad
\begin{minipage}{6cm}
\includegraphics[width=6cm]{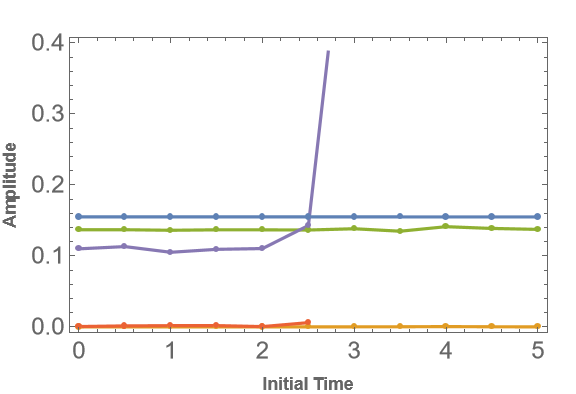}
\caption{\footnotesize Amplitude calculated for $n=8$ and $m=5$ with different initial time.}
\label{fig:3figsB}
\end{minipage}

\end{figure}
\noindent One would expect the amplitudes of the modes to depend on our initial value of $\psi$ but not on which part of the time domain we are fitting. However, for the generalised Bronnikov--Ellis wormhole, only the amplitudes associated with the fundamental modes and 2nd overtones (to some extent) are time-independent (Figs. \ref{fig:3figsA},\ref{fig:3figsB}). For almost all the other $\omega_{QNM}$, if we change the initial time of the fitting, the corresponding amplitudes remain stable up to some fitting domain, but beyond it, they begin to vary. Thus, we again fit the ringdown profiles with the modes with a stable amplitude 
\begin{figure}[H]
\centering
\parbox{7.5cm}{
\includegraphics[width=7.5cm]{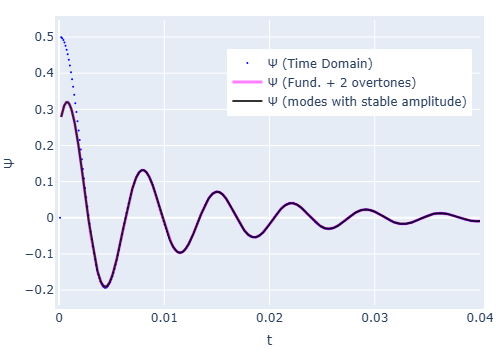}
\caption{\footnotesize Time-domain profile for all computed overtones and the modes with stable amplitude for $n=2$ and $m=5$.}
\label{1figsA}}
\qquad
\begin{minipage}{7.5cm}
\includegraphics[width=7.5cm]{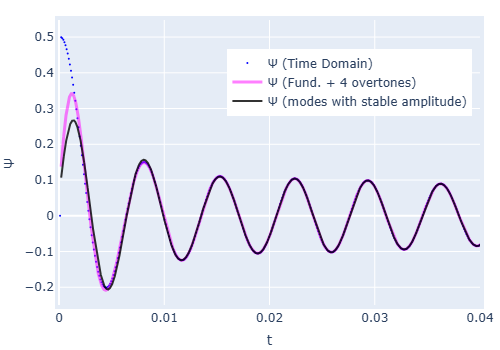}
\caption{\footnotesize Time-domain profile for all computed overtones and the modes with stable amplitude for $n=8$ and $m=5$.}
\label{1figsB}
\end{minipage}
\end{figure}

\noindent In both plots (Figs. \ref{1figsA}, \ref{1figsB}), the black line represents the contribution of the fundamental and 2nd overtone (the modes with stable amplitude), and the magenta line has the contribution of all the modes from tables \ref{tab:tab25} and \ref{tab:tab85}. For the $n=2$ case, the black and the magenta lines overlap (this is expected as we filtered out the 1st overtone with a very small amplitude). In the $n=8$ case, the two lines deviate at a very early time. The inclusion of all the modes gives a slightly better fit as the modes that were filtered out are also stable in the initial times, and their inclusion does not change the late-time behaviour, as these modes are damped quickly. Thus, 
we conclude that the inclusion of overtones can indeed help us achieve a better fit of the time-domain profile at earlier times.

\section{Redshift} \label{Redshift}

\noindent In the previous sections, we have computed the quasinormal modes for the massless generalised Bronnikov--Ellis wormhole. However, if the geometry under study is surrounded by a dark matter halo or any generic matter distribution, the QNMs calculated for the isolated compact object may differ from the observed values, as the environment may affect the generation of GW as well as their propagation. Such environments cause the spectrum to be red-shifted as documented in \cite{Barausse2014Can, pezzella2024quasinormal, Liu2025, Chen2023} for the black hole spacetime embedded in a halo. 

\noindent In the case of the Bronnikov--Ellis wormhole, similar redshifts in the fundamental mode can be found, if one considers the wormhole with non-zero ADM mass, as 
noted by the authors of \cite{blazquez2018scalar}. We will first find the QNMs (fundamental and overtones) for the $M\neq0$ case in the $n=2$ Bronnikov--Ellis geometry. Thereafter, in our second example, we will
find the QNMs for a
modified  $n=2$ geometry with a new redshift function 
having an extra parameter $M$ \cite{alfaro2024quasinormal}. 
In these two scenarios, using the QNMs obtained, we then
verify if we have a parameter-controlled redshift, similar to
the environment-induced redshift mentioned above. The modelling of such red-shifts is crucial as the red-shifted modes may appear to be QNMs of the original isolated geometry with a larger throat radius (as $\omega$ varies inversely with $b_0$).

\subsection{Bronnikov--Ellis geometry with asymptotic mass:} \label{AsymM}

\noindent Let us now see if redshift effects arise for the QNMs
of perturbations of Bronnikov--Ellis geometries. The ultrastatic Bronnikov--Ellis
geometry is not of use since it has only one parameter $b_0$ (throat radius)
which essentially appears as an overall scale factor in the QNMs. Thus, we 
either modify the shape function with extra parameters, in which case it is
no longer Bronnikov--Ellis or, alternatively, we may bring in the redshift function
which will have an extra parameter. We choose to work with this latter modification. 

\noindent The well-known, original line element for a Bronnikov--Ellis geometry with a redshift function is given as,
\begin{equation}
    ds^2 = -e^{f(l)}dt^2 + e^{-f(l)}[dl^2+(l^2+b_0^2)d\Omega_2^2]
\end{equation}
where, $f(l)=\frac{C}{b_0}\left(\arctan\left(\frac{l}{b_0}\right)-\frac{\pi}{2}\right)$. Note that due to the presence of $f(l)$, the energy density
as well as the pressures change, as discussed in Section \ref{Ellis_asymp_mass}.

\noindent We will work with this line element in order to demonstrate the redshift
effect in the QNMs.

\noindent The ADM mass ($M$) of this spacetime is related to the metric parameter $C$ as $C=2M$. One can plot the effective potentials for both scalar and axial perturbations numerically and observe that the symmetry in the potential barriers is lost with the introduction of non-zero asymptotic mass ($M$).

\begin{figure}[H]
\centering
\parbox{7.5cm}{
\includegraphics[width=7.5cm]{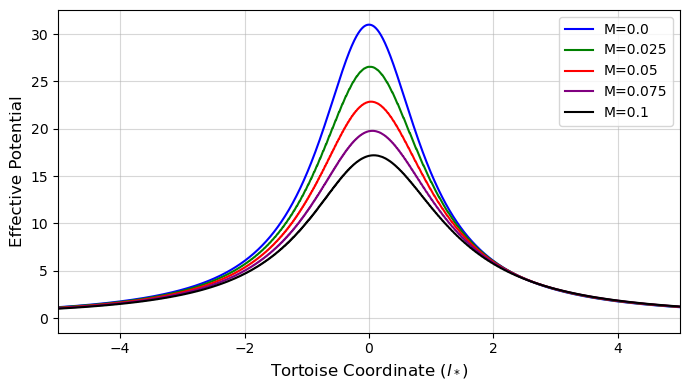}
\caption{\footnotesize Effective potentials (Scalar perturbation) for different values of M. (n = 2 and m = 5).}
\label{fig:Pot_Pro2_L5_sc}}
\qquad
\begin{minipage}{7.5cm}
\includegraphics[width=7.5cm]{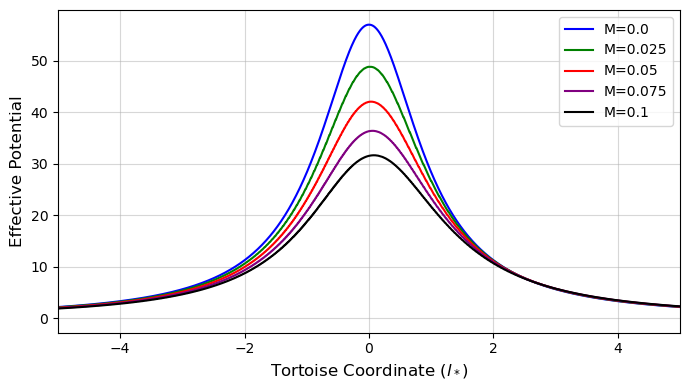}
\caption{\footnotesize Effective potentials (Scalar perturbation) for different values of M. (n = 2 and m = 7).}
\label{fig:Pot_Pro2_L7_sc}
\end{minipage}
\end{figure}

\begin{figure}[H]
\centering
\parbox{7.5cm}{
\includegraphics[width=7.5cm]{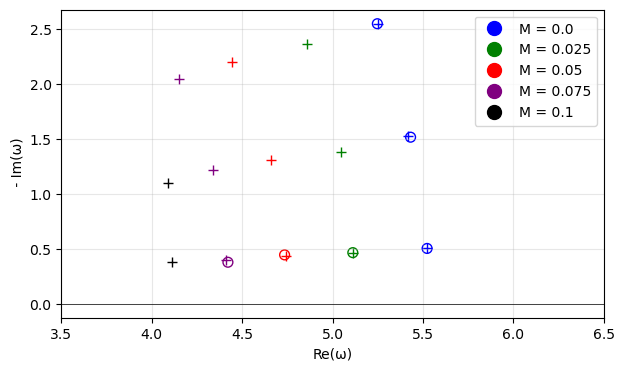}
\caption{\footnotesize QNMs (Scalar) for different values of M. (n = 2 and m = 5). The results of SM and Prony fit are represented by circles and plus signs, respectively.}
\label{fig:Mass_Pro2_L5_sc}}
\qquad
\begin{minipage}{7.5cm}
\includegraphics[width=7.5cm]{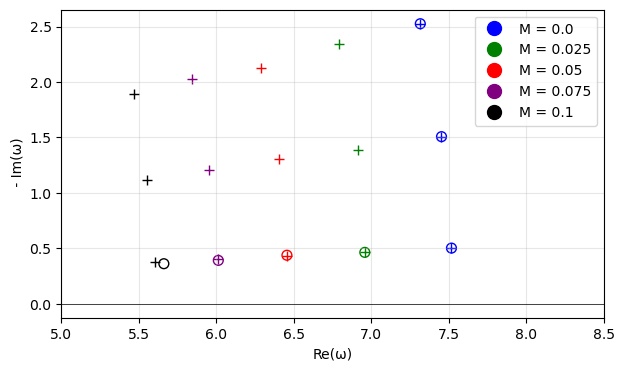}
\caption{\footnotesize QNMs (Scalar) for different values of M. (n = 2 and m = 7). The results of SM and Prony fit are represented by circles and plus signs, respectively.}
\label{fig:Mass_Pro2_L7_sc}
\end{minipage}
\end{figure}

\noindent The QNM spectra computed using the Prony fit (from $t=0$ to $t=40$ with $h=1/40$ using {\em Mathematica 14.2} for $M\neq 0$, $b_0=1$), for this geometry, are presented in the tables \ref{tab:qnm_values_Ellis_Atan_scalar},\ref{tab:qnm_values_Ellis_Atan_axial}
. However, in the figures we have presented some results from the spectral method.\\
\noindent The effective potential (scalar perturbation) for this metric can be expressed as Eq. \eqref{vsc}. Here, we have plotted the effective potentials for the $n=2$ case ($m=5$ in Fig. \ref{fig:Pot_Pro2_L5_sc} and $m=7$ in Fig. \ref{fig:Pot_Pro2_L7_sc}) and obtained single-peaked potentials with the maxima of the potential shifting to the right as we increase the value of $C$ ($=2M$). The QNMs corresponding to $m=5$ and $m=7$ are plotted in Figs. \ref{fig:Mass_Pro2_L5_sc} and \ref{fig:Mass_Pro2_L7_sc} respectively. \\

\begin{table}[H]
    \centering
    \footnotesize 
    \setlength{\tabcolsep}{4pt}  
    \renewcommand{\arraystretch}{1.2}   
    \singlespacing
    \begin{tabular}{|c|c|c|c|c|c|c|}
        \hline
        $m$ & Mode & $M=0b_0$  & $M=0.025b_0$  & $M=0.05b_0$ & $M=0.075b_0$  & $M=0.1b_0$\\
         \hline
        3 & 0 & $3.53443 - 0.50686i$ & $3.27166 - 0.46911i$ & $3.03576 - 0.43526i$ & $2.82383 - 0.40465i$  & $2.63255 - 0.37783i$\\
          
        \hline
        5 & 0 & $5.52282 - 0.50282i$ & $5.11243 - 0.46532i$ & $4.74378 - 0.43185i$ & $4.41259 - 0.40161i$  & $4.11427 - 0.37699i$\\
          & 1 & $5.41796 - 1.52258i$ & $5.04381 - 1.38243i$ & $4.65962 - 1.30864i$ & $4.33811 - 1.21722i$  & $4.08699 - 1.09940i$\\
          & 2 & $5.25123 - 2.54327i$ & $4.85855 - 2.36301i$ & $4.44292 - 2.19880i$ & $4.14809 - 2.04224i$  & $-$\\
          
        \hline
        7 & 0 & $7.51934 - 0.50106i$ & $6.95985 - 0.46386i$ & $6.45810 - 0.43037i$ & $6.00711 - 0.40048i$  & $5.60108 - 0.37349i$\\
          & 1 & $7.45576 - 1.50817i$ & $6.91398 - 1.38873i$ & $6.40556 - 1.30364i$ & $5.95396 - 1.20411i$  & $5.55057 - 1.11919i$\\
          & 2 & $7.31985 - 2.52412i$ & $6.79114 - 2.33889i$ & $6.28485 - 2.12915i$ & $5.84509 - 2.02687i$  & $5.46827 - 1.89265i$ \\
        \hline
        9 & 0 & $9.51561 - 0.50062i$ & $8.81008 - 0.46296i$ & $8.17473 - 0.42973i$ & $7.60378 - 0.39983i$  & $7.09023 - 0.37301i$\\
          & 1 & $9.43351 - 1.50092i$ & $8.76270 - 1.41073i$ & $8.13121 - 1.29187i$ & $7.55941 - 1.20308i$  & $7.04588 - 1.12225i$ \\
          & 2 & $9.35764 - 2.51354i$ & $8.64660 - 2.31498i$ & $8.02958 - 2.14977i$ & $7.50736 - 2.00358i$  & $6.98478 - 1.87115i$\\
        \hline
    \end{tabular}
    \caption{\footnotesize $\omega_{\text{QNM}}$ values for fundamental and overtones for Bronnikov--Ellis wormhole (Scalar perturbation) with $M=0,0.025b_0,0.05b_0,0.075b_0,0.1b_0$. (The results are obtained using Prony fit.) [The fundamental modes were computed in \cite{blazquez2018scalar}. We extended the analysis to include overtones.]}
    \label{tab:qnm_values_Ellis_Atan_scalar}
\end{table}

\begin{figure}[H]
\centering
\parbox{7.5cm}{
\includegraphics[width=7.5cm]{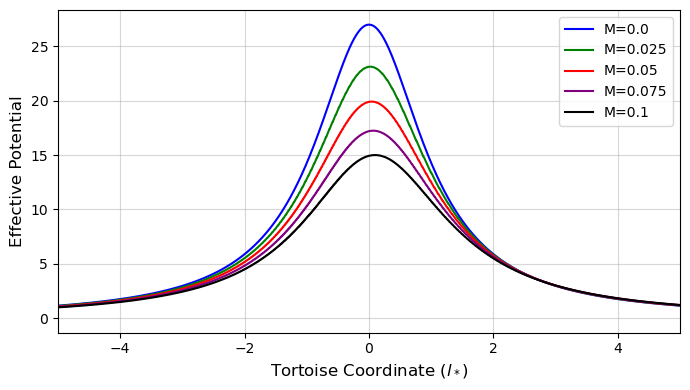}
\caption{\footnotesize Effective potentials (Axial Perturbation) for different values of M. (n = 2 and m = 5).}
\label{fig:Pot_Pro2_L5}}
\qquad
\begin{minipage}{7.5cm}
\includegraphics[width=7.5cm]{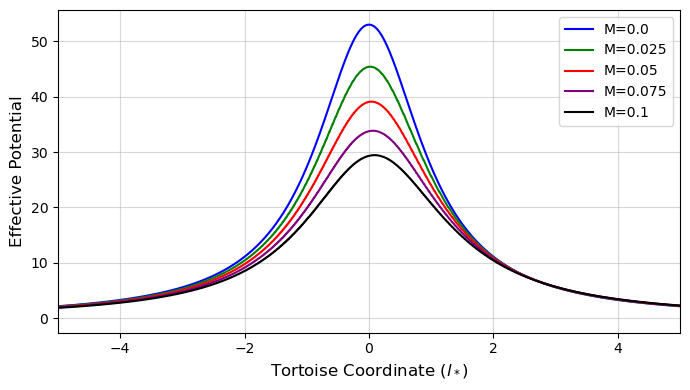}
\caption{\footnotesize Effective potentials (Axial Perturbation) for different values of M. (n = 2 and m = 7).}
\label{fig:Pot_Pro2_L7}
\end{minipage}
\end{figure}
\begin{figure}[H]
\centering
\parbox{7.5cm}{
\includegraphics[width=7.5cm]{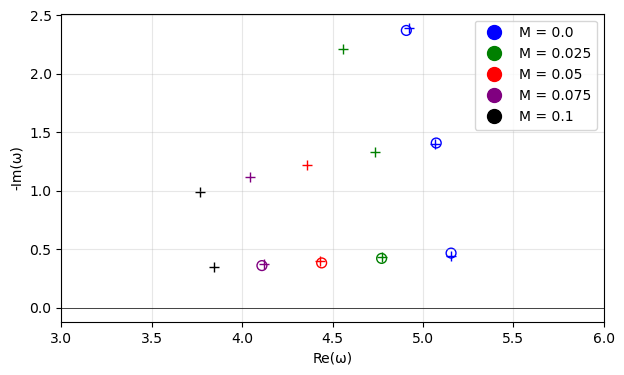}
\caption{\footnotesize QNMs (Axial Perturbation) for different values of M. (n = 2 and m = 5). The results of SM and Prony fit are represented by circles and plus signs, respectively.}
\label{fig:Mass_Pro2_L5}}
\qquad
\begin{minipage}{7.5cm}
\includegraphics[width=7.5cm]{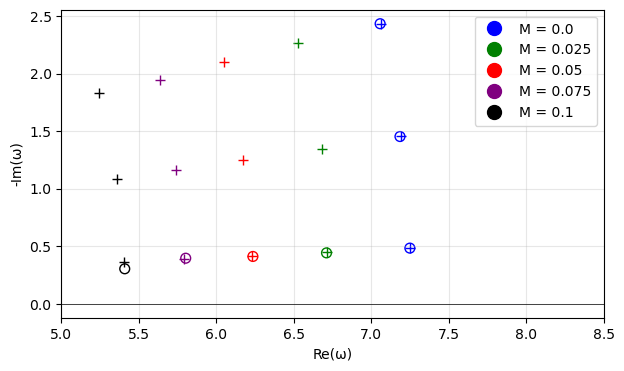}
\caption{\footnotesize QNMs (Axial Perturbation) for different values of M. (n = 2 and m = 7). The results of SM and Prony fit are represented by circles and plus signs, respectively.}
\label{fig:Mass_Pro2_L7}
\end{minipage}
\end{figure}

\noindent The effective potential for axial perturbation has the form Eq. \eqref{vax} and are plotted below. With increasing M ($=\frac{C}{2}$), the height of the barrier reduces and the potentials become more asymmetric (See Figs. \ref{fig:Pot_Pro2_L5}, \ref{fig:Pot_Pro2_L7}). The figures \ref{fig:Mass_Pro2_L5} and \ref{fig:Mass_Pro2_L7} display the QNM spectra for axial perturbation in $m=5$ and $m=7$ cases.

\noindent For this geometry, the increase of asymptotic mass (M) can be associated with redshift, i.e. for higher values of M, we have lower values of Re$(\omega)$. This phenomenon is similar to that observed by the authors of \cite{pezzella2024quasinormal}. We have presented the quasi-normal modes for small values of M, as we could not get overtones for higher values of M. The fundamental modes are in agreement with the results of \cite{blazquez2018scalar}.

\begin{table}[H]
    \centering
    \footnotesize  
    \setlength{\tabcolsep}{4pt}  
    \renewcommand{\arraystretch}{1.2}   
    \singlespacing
    \begin{tabular}{|c|c|c|c|c|c|c|}
        \hline
        $m$ & Mode & $M=0b_0$  & $M=0.025b_0$  & $M=0.05b_0$ & $M=0.075b_0$  & $M=0.1b_0$\\
         \hline
    
        5 & 0 & $5.15176 - 0.44366i$ & $4.77236 - 0.43237i$ & $4.42882 - 0.40144i$ & $4.12030 - 0.37367i$  & $3.84389 - 0.34924i$\\
          & 1 & $5.06686 - 1.39798i$ & $4.73155 - 1.33422i$ & $4.36034 - 1.21916i$ & $4.06004 - 1.12686i$  & $3.76455 - 0.99294i$\\
          & 2 & $4.91958 - 2.39447i$ & $4.55509 - 2.21569i$ & $-$ & $-$  & $-$\\
          
        \hline
        7 & 0 & $7.25073 - 0.48257i$ & $6.71107 - 0.44682i$ & $6.22750 - 0.41477i$ & $5.79355 - 0.38607i$  & $5.40297 - 0.36046i$\\
          & 1 & $7.19387 - 1.45787i$ & $6.68408 - 1.34773i$ & $6.17347 - 1.25075i$ & $5.74208 - 1.16371i$  & $5.35677 - 1.08268i$\\
          & 2 & $7.05993 - 2.43004i$ & $6.52963 - 2.26610i$ & $6.04865 - 2.09636i$ & $5.63629 - 1.94025i$  & $5.24520 - 1.82708i$ \\
        \hline
        9 & 0 & $9.30408 - 0.48927i$ & $8.61413 - 0.45250i$ & $7.99283 - 0.41976i$ & $7.43555 - 0.39076i$  & $6.97356 - 0.37212i$\\
          & 1 & $9.25255 - 1.47571i$ & $8.56364 - 1.36779i$ & $7.88056 - 1.26327i$ & $7.40840 - 1.17056i$  & $6.88158 - 1.09045i$ \\
          & 2 & $9.15068 - 2.45648i$ & $8.47519 - 2.27130i$ & $7.84263 - 2.09637i$ & $7.27180 - 1.96059i$  & $6.88756 - 1.86712i$\\
        \hline
    \end{tabular}
    \caption{\footnotesize $\omega_{\text{QNM}}$ values for fundamental and overtones for Bronnikov--Ellis wormhole (Axial perturbation) with $M=0,0.025b_0,0.05b_0,0.075b_0,0.1b_0$. (The results are obtained using Prony fit.) [The fundamental modes were computed in \cite{blazquez2018scalar}. We extended the analysis to include overtones.] }
    \label{tab:qnm_values_Ellis_Atan_axial}
\end{table}

\begin{figure}[H]
\centering
\parbox{7.5cm}{
\includegraphics[width=7.5cm]{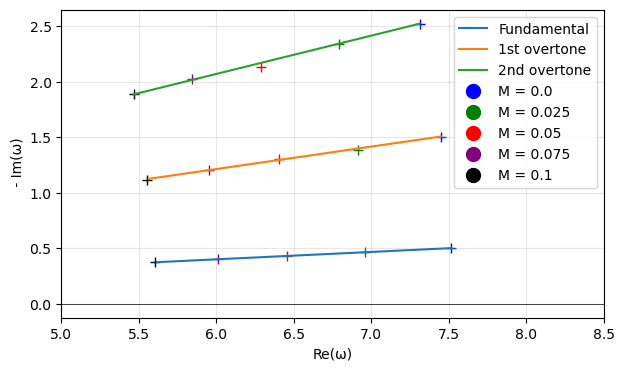}
\caption{Fit of the scalar $\omega_{QNM}$ ($m=7$) using the empirical fitting function.}
\label{fig:Ellis_atan_fit_scalar}}
\qquad
\begin{minipage}{7.5cm}
\includegraphics[width=7.5cm]{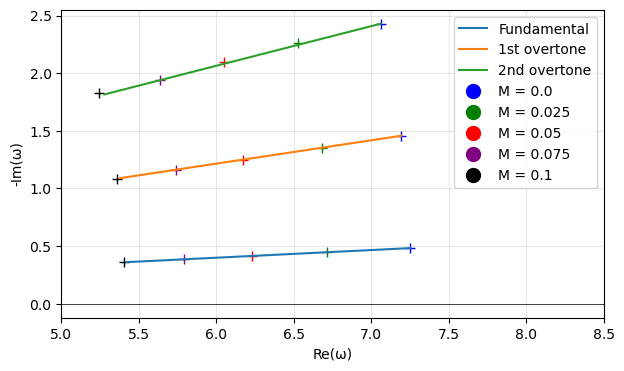}
\caption{Fit of the axial $\omega_{QNM}$ ($m=7$) using the empirical fitting function.}
\label{fig:Ellis_atan_fit_axial}
\end{minipage}
\end{figure}

\noindent It is interesting to note that both the real and imaginary parts of the $\omega_{QNM}$ decrease with increasing M, with the fundamental mode and the overtones following the same pattern. We can estimate the QNMs for a certain M value from the spectra of the original Bronnikov--Ellis wormhole ($M=0$ case) using the purely empirical formula
\begin{equation*}
    \omega_M=\omega_0 exp\left(-\frac{\pi M}{1+0.8M}\right)
\end{equation*}

\noindent where, $\omega_0$ is quasi-normal mode corresponding to $M=0$ case and $\omega_M$ denotes quasi-normal modes corresponding to $M\neq0$ cases. This relation is true for both scalar and axial perturbations of the $n=2$ geometry and holds for all the values of the overtone number and the angular momentum parameter (i.e. the ones we have obtained numerically). We are not aware of the origin of this empirical relation,
though, for small values of the exponent, an approximate linear relation seems to work, as evident from the plots.\\
\noindent In Figs. \ref{fig:Ellis_atan_fit_scalar} (scalar) and \ref{fig:Ellis_atan_fit_axial} (axial), we have plotted the QNMs (obtained using the Prony Fit method) and the above empirical fitting function for the $n=2$ and $m=7$ case. 

\noindent For the $n=2$ case, one can observe that the increasing ADM mass ($M$) has a very similar effect on fundamental mode and overtones. Due to this feature of the QNM spectra, the ringdown signal of a Bronnikov-Ellis wormhole with non-zero $M$ can be mistaken as that of the massless Bronnikov-Ellis wormhole with a larger throat radius ($b_0$).

\subsection{Bronnikov--Ellis geometry with a chosen redshift function:} \label{Mod_Ellis}

\noindent Apart from the geometry we studied in the previous subsection, we may also consider the Bronnikov--Ellis wormhole (the $n=2$ case) with a different non-zero redshift function, introduced and discussed earlier in Section \ref{chosen_redshift_func}. The line element is given as,

\begin{equation*}
    ds^2 = - \left(1-\frac{2M}{r}\right)dt^2 + \frac{dr^2}{1-\frac{b_0^2}{r^2}} + r^2 d\theta^2 + r^2 sin^2\theta d\varphi^2
\end{equation*}
Here, as stated before, the energy density does not change on varying $M$,
but the pressures change.
We will now see if a redshift effect of the QNMs is present when we tune the value of $M$.

\noindent The Schrodinger-like equation for perturbations of this spacetime is
\begin{equation}
    \frac{d^2\psi(r)}{dr_*^2}+(\omega^2-V_{eff}(r))\psi(r)=0 \label{Master_redshift}
\end{equation}
where the tortoise coordinate $r_*$ has the form 

\begin{align*}
    r_*= &\frac{1}{ \sqrt{r \left(r^2-b_0^2\right) (r-2 M)}(4 M^2-b_0^2)}\bigg(b_0 \bigg(b_0^2 k_1 (r-2 M)^2 \left(E\left(k_0|a_1\right)-F\left(k_0|a_1\right)\right)+2 b_0 M \big(k_1 (r-2 M)^2 \big(E\left(k_0|a_1\right)\\
    &-\Pi \left(a_0;k_0|a_1\right)\big)-2 M r\big)-4 k_1 M^2 (r-2 M)^2 \left(F\left(k_0|a_1\right)-\Pi \left(a_0;k_0|a_1\right)\right)+b_0^3 r\bigg)+r^3(4 M^2-b_0^2)\bigg)
\end{align*}

\noindent with $a_0=\frac{2 b_0}{b_0+2 M}$, $a_1=\frac{4 M}{b_0+2 M}$, $k_0=sin^{-1}\sqrt{\frac{\left(b_0+2 M\right) \left(r-b_0\right)}{2 b_0 (r-2 M)}}$, $k_1= \sqrt{\frac{r \left(b_0-2 M\right)\left(b_0^2-4 M^2\right) \left(r^2-b_0^2\right)}{b_0^3 (r-2 M)^3}}$, $E(...)=EllipticE$, $F(...)=EllipticF$ and $\Pi(...)=EllipticPi$ function.\\
\noindent Now, to compute the QNMs with the spectral method, it is convenient to consider two separate ansatz \cite{alfaro2024quasinormal} (one for odd overtones and one for even).
\begin{equation*}
    \psi(r)_{even}=e^{i\omega r}r^{iM\omega}\phi(r)_{even} \hspace{2cm} \psi(r)_{odd}=(r-b_0)^\frac{1}{2}e^{i\omega r}r^{iM\omega-\frac{1}{2}}\phi(r)_{odd}
\end{equation*}
Further, we have used the transformation $x=1-\frac{2b_0}{r
}$ to map the radial coordinate $r\in(b_0,\infty)$ to $x\in[-1,1]$ \cite{batic2024unified}. This procedure is applicable for both scalar and axial perturbations. The tables below (\ref{tab:qnm_values_Red_scalar1}, \ref{tab:qnm_values_Red_Axial1}) contain results obtained using the spectral method. However, we have also shown the results from the Prony fit method in the plots.

\noindent For scalar perturbation the expression for $V_{eff}$ is
\begin{equation*}
    V_{sc}(r)=\frac{b_0^2 (r-3 M)+r^2 \left(\left(-2 m^2-2 m+1\right) M+m (m+1) r\right)}{r^5}
\end{equation*}

\begin{figure}[H]
\centering
\parbox{7.5cm}{
\includegraphics[width=7.5cm]{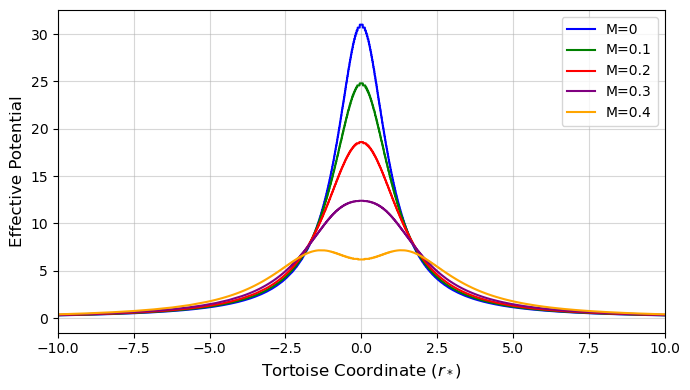}
\caption{\footnotesize Effective potentials (Scalar Perturbation) for different values of $M$. (n = 2 and m = 5).}
\label{fig:Pot_Pro1_L5_sc}}
\qquad
\begin{minipage}{7.5cm}
\includegraphics[width=7.5cm]{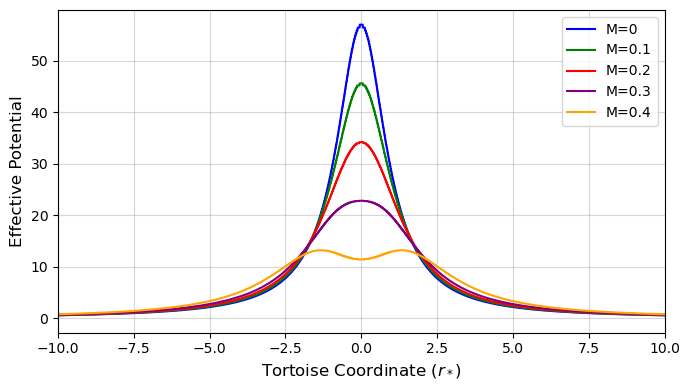}
\caption{\footnotesize Effective potentials (Scalar Perturbation) for different values of $M$. (n = 2 and m = 7).}
\label{fig:Pot_Pro1_L7_sc}
\end{minipage}
\end{figure}

\noindent As shown in the figures below (Figs. \ref{fig:Pot_Pro1_L5_sc}, \ref{fig:Pot_Pro1_L7_sc}), the effective potential changes from being a single barrier to a double barrier for higher $M$. The potentials have been plotted numerically.

\begin{figure}[H]
\centering
\parbox{7.5cm}{
\includegraphics[width=7.5cm]{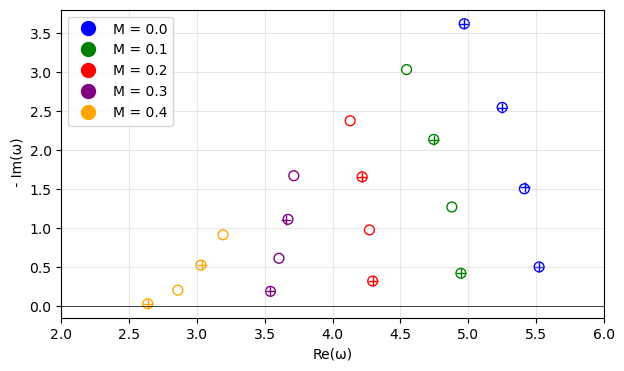}
\caption{\footnotesize QNMs (Scalar) for different values of $M$. (n = 2 and m = 5). The results of SM and Prony fit are represented by circles and plus signs, respectively.}
\label{fig:Mass_Pro1_L5_sc}}
\qquad
\begin{minipage}{7.5cm}
\includegraphics[width=7.5cm]{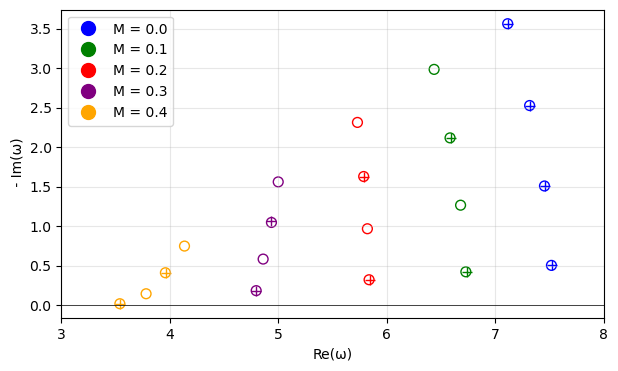}
\caption{\footnotesize QNMs (Scalar) for different values of $M$. (n = 2 and m = 7). The results of SM and Prony fit are represented by circles and plus signs, respectively.}
\label{fig:Mass_Pro1_L7_sc}
\end{minipage}
\end{figure}

\noindent For this geometry, as we increase the value of $M$, both the real and imaginary parts of the QNM spectra decrease (See Figs. \ref{fig:Mass_Pro1_L5_sc} ,\ref{fig:Mass_Pro1_L7_sc}). Thus, an increase in $M$ causes a redshift. On careful observation, one notices that for lower values of $M$, the real parts of quasinormal modes decrease with the overtone number.  However, this pattern reverses when we go to higher $M$ values. The height of the effective potential decreases as we keep on increasing $M$, and after a certain value, the effective potential becomes double-peaked. This may explain the above feature of the quasi-normal spectra.\\
For a fixed overtone number, the $\omega_{QNM}$ of scalar perturbation corresponding to different $M$ values can be roughly estimated by linear fits. The fitting parameter will vary with the overtone number and angular momentum parameter (m).

\begin{table}[H]
    \centering
    \footnotesize  
    \setlength{\tabcolsep}{4pt}  
    \renewcommand{\arraystretch}{1.2}   
    \singlespacing
    \begin{tabular}{|c|c|c|c|c|c|c|}
        \hline

        $m$ & Mode & $M=0b_0$  & $M=0.1b_0$  & $M=0.2b_0$ & $M=0.3b_0$  & $M=0.4b_0$\\
         \hline

        \hline
        1 & 0 & $1.57271 - 0.52970i$ & $1.42443 - 0.44987i$ & $1.26184 - 0.35838i$ & $1.07980-0.25189i$  & $0.86152-0.12862i$\\
          & 1 & $1.25957 - 1.70681i$ & $1.18531 - 1.43536i$ & $1.12961 - 1.14815i$ & $1.08147-0.84604i$  & $1.01104-0.51803i$\\
          & 2 & $-$ & $-$ & $0.97385 - 2.14007i$ & $1.09839-1.63586i$  & $1.18106-1.10171i$ \\

        \hline
        3 & 0 & $3.53429 - 0.50692i$ & $3.17040-0.42628i$ & $2.76272-0.32941i$ & $2.29567-0.20688i$  & $1.73832-0.06152i$\\
          & 1 & $3.39029 - 1.51877i$ & $3.06406-1.29199i$ & $2.71576-1.00573i$ & $2.34754-0.67242i$  & $1.93405-0.30467i$\\
          & 2 & $3.10063 - 2.62534i$ & $2.85146-2.20091i$ & $2.61615-1.73263i$ & $2.38730-1.23570i$  & $2.08989-0.70557i$ \\
          & 3 & $2.71863 - 3.55029i$ & $-$ & $2.46305-2.54081i$ & $2.39244-1.88891i$  & $2.23465-1.18172i$ \\
          
        \hline
        5 & 0 & $5.52231 - 0.50296i$ & $4.94551-0.42182i$ & $4.29517-0.32260i$ & $3.54142-0.19058i$  & $2.63686-0.03204i$\\
          & 1 & $5.41372 - 1.50367i$ & $4.87885-1.27072i$ & $4.27080-0.97686i$ & $3.60420-0.61373i$  & $2.85899-0.20555i$\\
          & 2 & $5.25011 - 2.54570i$ & $4.74524-2.13600i$ & $4.21729-1.65629i$ & $3.67073-1.11220i$  & $3.02866-0.52636i$ \\
          & 3 & $4.97014 - 3.61705i$ & $4.54492-3.03031i$ & $4.12908-2.37426i$ & $3.71354-1.67104i$  & $3.19144-0.91448i$ \\
          
        \hline
        7 & 0 & $7.51649 - 0.50162i$ &$6.72760-0.42027i$  & $5.83575-0.31996i$ & $4.79524-0.18192i$  & $3.53888-0.01635i$\\
          & 1 & $7.45225 - 1.50671i$ & $6.67910-1.26365i$ & $5.82019-0.96550i$ & $4.85928-0.58145i$  & $3.78057-0.14253i$ \\
          & 2 & $7.31542 - 2.52531i$ & $6.58191-2.11572i$ & $5.78590-1.62677i$ & $4.93587-1.04502i$  & $3.95845-0.40847i$\\
          & 3 & $7.11330 - 3.56071i$ & $6.43513-2.98262i$ & $5.72857-2.31139i$ & $4.99835-1.55865i$  & $4.13449-0.74607i$\\
        \hline
    \end{tabular}
    \caption{\footnotesize $\omega_{\text{QNM}}$ values for fundamental and overtones for Bronnikov--Ellis wormhole (Scalar perturbation) with $M=0,0.1b_0$, $0.2b_0$, $0.3b_0$ and $0.4b_0$.  [The QNM spectra for this geometry (single-peaked scalar potentials) were published in \cite{alfaro2024quasinormal}. We extended this to double-peaked potentials (Higher $M$)]. [Results obtained using the spectral method ($b_0=1$).] }
    \label{tab:qnm_values_Red_scalar1}
\end{table}

\begin{figure}[H]
\centering
\parbox{7.5cm}{
\includegraphics[width=7.5cm]{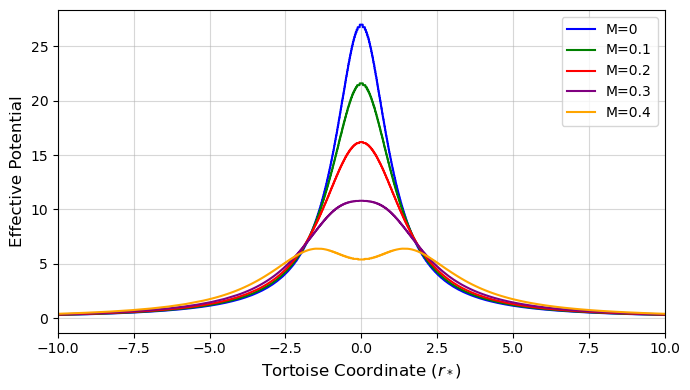}
\caption{\footnotesize Effective potentials (Axial Perturbation) for different values of $M$. (n = 2 and m = 5).}
\label{fig:Pot_Pro1_L5}}
\qquad
\begin{minipage}{7.5cm}
\includegraphics[width=7.5cm]{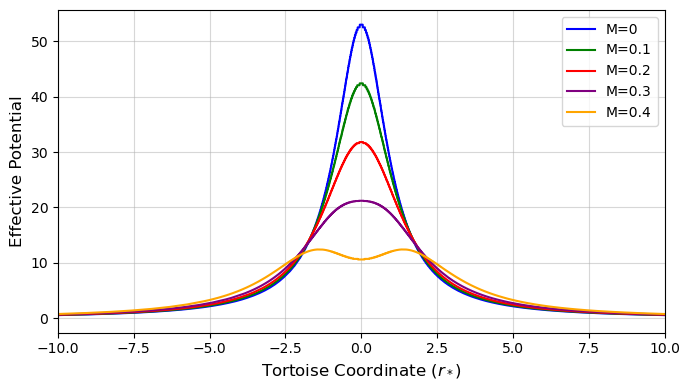}
\caption{\footnotesize Effective potentials (Axial Perturbation) for different values of $M$. (n = 2 and m = 7).}
\label{fig:Pot_Pro1_L7}
\end{minipage}
\end{figure}

\noindent In the axial perturbation case, the effective potential in Eq. \eqref{Master_redshift} has the form
\begin{equation*}
    V_{ax}(r)=\frac{b_0^2 (7 M-3 r)+r^2 (m (m+1) r-(2 m (m+1)+1) M)}{r^5}
\end{equation*}
\noindent Similar to the scalar perturbation, the effective potential due to axial perturbation changes from being a single barrier to a double barrier for higher $M$ (See Figs. \ref{fig:Pot_Pro1_L5} ,\ref{fig:Pot_Pro1_L7}).

\begin{figure}[H]
\centering
\parbox{7.5cm}{
\includegraphics[width=7.5cm]{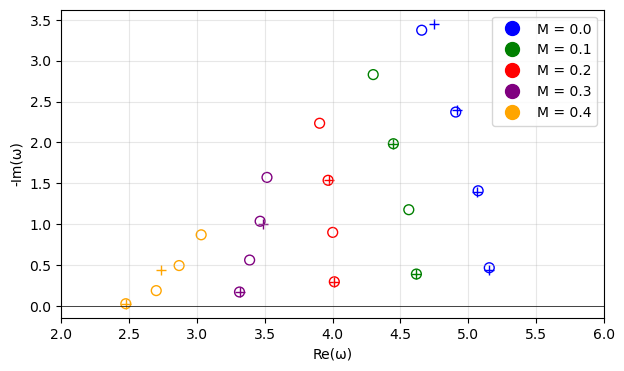}
\caption{\footnotesize QNMs (Axial) for different values of $M$. (n = 2 and m = 5). The results of SM and Prony fit are represented by circles and plus signs, respectively.}
\label{fig:Mass_Pro1_L5}}
\qquad
\begin{minipage}{7.5cm}
\includegraphics[width=7.5cm]{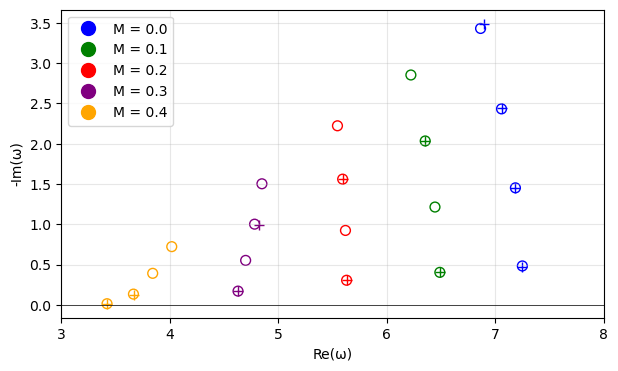}
\caption{\footnotesize QNMs (Axial) for different values of $M$. (n = 2 and m = 7). The results of SM and Prony fit are represented by circles and plus signs, respectively.}
\label{fig:Mass_Pro1_L7}
\end{minipage}
\end{figure}

\noindent The quasi-normal spectra for axial perturbation have a similar redshift effect as the QNMs for scalar perturbation (Figs. \ref{fig:Mass_Pro1_L5} ,\ref{fig:Mass_Pro1_L7}). However, unlike the scalar perturbation case, here, a linear fit does not estimate the $\omega_{QNM}$ of the same overtone order.

\begin{table}[H]
    \centering
    \footnotesize  
    \setlength{\tabcolsep}{4pt}  
    \renewcommand{\arraystretch}{1.2}   
    \singlespacing
    \begin{tabular}{|c|c|c|c|c|c|c|}
        \hline

        $m$ & Mode & $M=0b_0$ & $M=0.1b_0$  & $M=0.2b_0$ & $M=0.3b_0$  & $M=0.4b_0$\\
         \hline

        \hline
        3 & 0 & $2.95241-0.40998i$ & $2.65202-0.34116i$ & $2.31704-0.25890i$ & $1.93583-0.15765i$  & $1.47895-0.04526i$\\
          & 1 & $2.86643-1.25767i$ & $2.59943-1.05329i$ & $2.32019-0.81701i$ & $2.02320-0.54789i$  & $1.68143-0.25349i$\\
          & 2 & $2.66056-2.19085i$ & $2.47232-1.84041i$ & $2.28366-1.46458i$ & $2.09262-1.06263i$  & $1.83772-0.62125i$ \\

        \hline
        5 & 0 & $5.15482-0.46725i$ & $4.61743-0.39009i$ & $4.01226-0.29556i$ & $3.31317-0.17059i$  & $2.47482-0.02713i$\\
          & 1 & $5.07286-1.40857i$ & $4.56144-1.17693i$ & $4.00038-0.89951i$ & $3.38786-0.56119i$  & $2.69934-0.18747i$\\
          & 2 & $4.90746-2.37118i$ & $4.44773-1.98388i$ & $3.96646-1.53674i$ & $3.46529-1.03582i$  & $2.86794-0.49395i$ \\
          & 3 & $4.65661-3.37224i$ & $4.27300-2.82560i$ & $3.89988-2.21918i$ & $3.51728-1.57299i$  & $3.03059-0.87000i$ \\
          
        \hline
        
        7 & 0 & $7.24818-0.48308i$ & $6.48787-0.40372i$ & $5.62868-0.30567i$ & $4.62788-0.17089i$  & $3.42055-0.01444i$\\
          & 1 & $7.18438-1.45261i$ & $6.44308-1.21430i$ & $5.61809-0.92386i$ & $4.69804-0.55209i$  & $3.66405-0.13378i$\\
          & 2 & $7.05634-2.43256i$ & $6.35314-2.03436i$ & $5.59222-1.56056i$ & $4.78041-1.00184i$  & $3.84139-0.39161i$ \\
          & 3 & $6.86588-3.43084i$ & $6.21733-2.87112i$ & $5.54494-2.22360i$ & $4.84772-1.50282i$  & $4.01693-0.72270i$ \\
          
        \hline
        9 & 0 & $9.30171-0.48964i$ & $8.32354-0.40939i$ & $7.21638-0.30982i$ & $5.92045-0.16947i$  & $4.34739-0.00712i$\\
          & 1 & $9.25035-1.47100i$ & $8.28712-1.23004i$ & $7.20768-0.93382i$ & $5.98667-0.54158i$  & $4.60839-0.09362i$ \\
          & 2 & $9.14741-2.45870i$ & $8.21414-2.05635i$ & $7.18761-1.56991i$ & $6.07049-0.97279i$  & $4.79271-0.31009i$\\
          & 3 & $8.99239-3.45726i$ & $8.10424-2.89222i$ & $7.15227-2.22372i$ & $6.14625-1.44932i$  & $4.97551-0.60743i$\\
        \hline
    \end{tabular}
    \caption{\footnotesize $\omega_{\text{QNM}}$ values for fundamental and overtones for Bronnikov--Ellis wormhole (Axial perturbation) with $M=0,0.1b_0$, $0.2b_0$, $0.3b_0$ and $0.4b_0$.  [The QNM spectra for this geometry (single-peaked scalar potentials) were published in \cite{alfaro2024quasinormal}. We extended this to axial perturbation and double-peaked potentials (Higher $M$)]. [Results obtained using the spectral method ($b_0=1$).] }
    \label{tab:qnm_values_Red_Axial1}
\end{table}
\noindent For this modified Bronnikov-Ellis wormhole, the spectra change with increasing $M$. For the $M=0$ case, the real part decreases with increasing overtone number, but this trend gradually reverses with increasing $M$. Though both Re$(\omega)$ and $|Im(\omega)|$ decrease with $M$ (similar effect as increasing $b_0$ of the original Bronnikov-Ellis spacetime), the detection of the first few overtones can distinguish this geometry (with non-zero redshift function) from the original Bronnikov-Ellis wormhole.

\section{Remarks and future directions} \label{conclusion}

\noindent In our work here, we have revisited the problem of computing the QNMs of the generalised Bronnikov-Ellis wormholes with emphasis on the overtones.
Further, we have also studied the `redshift' of QNMs induced by
varying the metric parameters 
in analogy with the environment-induced proposed recently. Below, we briefly summarise our main results.

\

$\bullet$ We extend the ultrastatic, generalised (two parameter--$b_0$, $n>2$) Bronnikov-Ellis wormhole formulated in \cite{kar1995resonances}  by constructing a modified metric with a non-zero asymptotic mass $M$ (furthering  the analysis in \cite{blazquez2018scalar} for $n=2$). The geometry, energy-momentum tensor and the energy conditions for the
new three-parameter ($b_0$, $n$, $M$) space-times are discussed briefly, followed by the 
computation of the corresponding effective potentials for scalar and axial gravitational perturbations.

$\bullet$ We have used the spectral method (frequency domain) and the Prony Fit (time domain) to compute the QNM spectra (scalar and axial gravitational) for the ultrastatic case. Beyond the fundamental modes, we have computed 2-3 overtones (4 in some cases), 
which, as we show, may help in distinguishing different geometries (i.e. with different $n$). We have also pointed out key features of the QNM spectra and how the shape of the effective potentials may explain the
trends in QNM behaviour.

$\bullet$ The range of the throat radius ($b_0$) for detectable QNMs (in aLIGO) in the ultrastatic case is found.  We also establish, for a few cases, using the
time-domain waveform, the crucial role that overtones may play in fitting the 
signal at early times (when SNR is high).

$\bullet$ Finally, we focused on the $n=2$ wormhole family with asymptotic mass $M$ and observed how the increase in $M$ causes a systematic redshift (lowering of values of Re$(\omega)$). We further observed this redshift effect (w.r.t. changing $M$) for the original Bronnikov--Ellis wormhole but with a chosen redshift function ($e^{2 \Psi(r)}=1-\frac{2M}{r}$) \cite{alfaro2024quasinormal}. The distinguishability of these geometries from their massless counterparts (with larger $b_0$) has also been 
mentioned.

\

\noindent Various extensions of this work here are possible. 
We have not really succeeded in obtaining a large number of overtones
which can give us a clearer picture of the full quasinormal mode spectrum.
This aspect surely needs to be handled better. 
Further, though we have identified distinct features and have tried to
demonstrate how overtones may affect a signal, a better and more
complete understanding is desirable. In addition, learning more
about polar perturbations and their overtones remains a difficult task
which needs to be addressed.
Finally, in the context of the redshift, though we see a correlation
with the behaviour of the effective potentials, the results are
case-specific and need to be understood in a more generic way.

\noindent We hope to return to the abovementioned issues in future
investigations.

\noindent \section*{ACKNOWLEDGEMENTS}

\noindent ABM thanks the Indian Institute of Technology Kharagpur, India, for supporting
him through a fellowship and allowing access to available computational facilities.

\bibliographystyle{apsrev} 
\bibliography{ref}

\newpage

\appendix

\section{Methods used to find the QNMs} \label{method}

\noindent Computation of quasinormal modes requires solving linearised perturbation equations of spacetime under specific boundary conditions (purely outgoing at spatial infinities of the tortoise coordinate). These equations are, in general, non-trivial to solve analytically, and so we employ a variety of numerical, analytical and semi-analytical methods. Below we provide short summaries of the key methods we have used here to determine the QNMs of the generalised Bronnikov--Ellis wormholes.

\subsection {Time domain and Prony fit} \label{Prony}

\noindent The time domain profile represents the temporal evolution of the field $\psi$ in a fixed background geometry. It was originally formulated in \cite{gundlach1994late}. In section \ref{eff}, we derived the master wave equation, which may also be rewritten without the separability ansatz for the field. Thus, the equation, using the new coordinates $u$ and $v$, such that $du = dt - dx, \hspace{0.2cm} dv = dt + dx$, reads \cite{konoplya2011quasinormal}:
\begin{equation}
4\frac{\partial^2 \psi}{\partial u \partial v} + V(u, v) \psi = 0.
\end{equation}

\noindent Rewriting the time evolution operator in the new coordinates and discretising the wave equation, we obtain,
\begin{equation}
\psi(N) = \psi(W) + \psi(E) - \psi(S) - \frac{h^2}{8} V(S) [\psi(W) + \psi(E)] + O(h^4), \label{td1}
\end{equation}
where \( S = (u, v) \), \( W = (u + h, v) \), \( E = (u, v + h) \), and \( N = (u + h, v + h) \) and $h$ is the integration step size for both $u$ and $v$. Eqn. \eqref{td1} enables us to compute \( \psi \) iteratively on the $u-v$ grid based on the initial data specified on null surfaces \( u = u_0 \) and \( v = v_0 \). The values of $\psi(N)$,  thus calculated, give us the time-domain profile for the given potential.\\
\noindent The time-domain profile of \( \psi \) is then fitted using the Prony method $\psi(t) \approx \sum_{j=1}^p C_j e^{-i \omega_j t}$. Here \( \omega_j \) are the quasinormal frequencies and $C_j$ are the corresponding amplitudes. We assume that the ringdown starts at $t=0$ and ends at $t=N_1h$, and approximate the profile data as a sum of $p \hspace{0.1cm} (\leq \frac{N_1+1}{2})$ terms in the following way:
\begin{equation}
x_n \approx \psi(nh) = \sum_{j=1}^p C_j z_j^n, \hspace{2cm} z_j=e^{-i\omega_j h}
\end{equation}

\noindent In the Prony fit method, it is assumed that the $\psi(nh)$ satisfies the recurrence relation
\begin{equation}
    \sum_{m=0}^p a_m \psi((n-m)h)=0 \hspace{0.2cm} \implies \hspace{0.2cm} \sum_{m=0}^p a_m x_{n-m}=0 \label{p1}
\end{equation}
where $a_0=1$. This leads to the matrix equation:

\begin{equation}
\begin{bmatrix}
x_{p-1} & x_{p-2} & \hdots & x_{0} \\
x_{p} & x_{p-1} & \hdots & x_{1}  \\
\vdots & \vdots & \ddots &  \vdots  \\
x_{N_1-1} & x_{N_1-2} & \hdots & x_{N_1-p} 
\end{bmatrix}
\begin{bmatrix}
a_1 \\
a_2 \\
\vdots \\
a_p
\end{bmatrix}
=
-\begin{bmatrix}
x_{p} \\
x_{p+1} \\
\vdots \\
x_{N_1}
\end{bmatrix}
\hspace{1cm} \implies
X \cdot a = -x, \label{p2}
\end{equation}
Eqn. \eqref{p2} is solved to get $a= -(X^\dagger X)^{-1} X^\dagger x$,
where \( X^\dagger \) denotes the Hermitian transpose of \( X \). 
This yields all the coefficients $a_m$, and thus we may form the master polynomial from Eqn. \eqref{p1}. This polynomial can be solved numerically to obtain the roots $z_j$, and subsequently, the QNMs are calculated from these roots
\begin{equation}
\omega_j = i \frac{\ln(z_j)}{h}.
\end{equation}

\noindent Having computed the quasi-normal frequencies, we can also obtain their corresponding amplitudes using the Prony fit. For this, we need to define a $(p\times N_1)$ matrix $A_{ij}=z_{i-1}^{j}$ and a vector $v=[x_1,x_2,\hdots,x_{N_1}]$.
We get the amplitudes by calculating $C=[C_1,C_2,\hdots,C_p]= (A^\dagger A)^{-1} A^\dagger v$. These amplitudes are generally complex numbers; however, we have reported only the absolute values in the appendix \ref{Table}.

\noindent In the method discussed above, we get $p$-number of solutions, but most of them are numerical artefacts and not actual QNMs of the system. To mitigate this issue, we can repeat the above steps with different values of $h$ and $N_1$ and accept only the results common to both sets of solutions (with appropriate tolerance) as the actual QNMs.\\
The time-domain and Prony fit method works well for most effective potentials which do not have any discontinuities. But this method often fails to return the $\omega_{QNM}$ with low relative amplitude (we notice this feature for odd overtones in Sec. \ref{result}).

\subsection {The spectral method}

\noindent Using the corresponding equation for a
perturbation (scalar or axial gravitational), another method
known as the spectral method, may also be employed to find the QNMs of the system. In our analysis, we will use the Chebyshev grid and thus we have to map our coordinate to a finite interval, i.e. $[-1,1]$. The tortoise coordinate $l\in (-\infty, +\infty)$ may be mapped to $x\in [-1,1]$ using the transformation $x=\frac{2}{\pi}\arctan(l)$ \cite{batic2024unified}.

\noindent For the generalised Bronnikov--Ellis wormhole with $M=0$, we write the Schrodinger-like equation in terms of the tortoise coordinate ($l$),
\begin{equation}
     \frac{d^2 \psi}{d l^2} +(\omega^2- V(l)) \psi = 0
\end{equation}
Now, we need to ensure that the `purely outgoing' boundary conditions at both infinities of the tortoise coordinate. For that, we examine the behaviour of $\psi$ near the throat and at infinity and define a new variable $\phi(x)$ using a suitable ansatz: \cite{batic2024unified}.
\begin{equation}
    \psi(x)=exp\left(i\omega \frac{2l}{\pi}\arctan(l)\right)\phi(x) \label{ansatz}
\end{equation}

\noindent Next, the function $\phi$ is discretized  with the Chebyshev polynomial of first kind $C_i(x)$ as  \cite{jansen2017overdamped}
\begin{equation}
    \phi(x) \approx \sum_{i=0}^N \phi(x_i)C_i(x)
\end{equation}
with $x_i$ being the collocation points (Chebyshev roots) given by $x_i=\frac{(2i+1)\pi}{2(N+1)}$ $(i = {0,1,2,...,N})$.

\noindent Thus, we can express the master equation in the form 
\begin{equation}
c_2(x,\omega)\phi''(x)+c_1(x,\omega)\phi'(x)+c_0(x,\omega)\phi(x)=0 \label{sp1}
\end{equation}
and define $c_{p,0}$ as part of $c_0$ with coefficient of $\omega^p$, $c_{q,2}$ as part of $c_2$ with coefficient of $\omega^q$ and so on \cite{jansen2017overdamped}. We compute these expressions and $\phi$ at the co-location points and get numerical matrices of the form
\begin{equation}
(M_a)_{ij}=c_{0,a}D^{(0)}_{ij}+c_{1,a}D^{(1)}_{ij}+c_{2,a}D^{(2)}_{ij} \hspace{1cm} a = {1,2,...} \label{c_aa}
\end{equation}

\noindent where, $D^{(n)}_{ij}$ represents n-th derivative of $C_i(x_j)$. If the highest power of $\omega$ in Eqn. \eqref{sp1} is $p$, then we will need to compute $p$ such matrices. Finally, \eqref{sp1} is expressed as 
\begin{equation}
    (M_0+\omega M_1+\omega^2M_2+...+\omega^pM_p)\phi=0 \label{sp2}
\end{equation}
Solving the matrix equation \eqref{sp2} will return the quasi-normal frequencies of the system.\\

\noindent We can consider the example of scalar perturbation in the ultra-static Bronnikov--Ellis geometry (Eqn. \eqref{geo4}), where, after using the ansatz (Eqn. \eqref{ansatz}) we have

\begin{align*}
     c_0= &-m (m+1) \left(\tan ^n\left(\frac{\pi  x}{2}\right)+1\right)^{-2/n}-\frac{(n-1) \tan ^{n-2}\left(\frac{\pi  x}{2}\right)}{\left(\tan ^n\left(\frac{\pi  x}{2}\right)+1\right)^2}-\left(x^2-1\right) \omega ^2\\
     &+\frac{2 i \pi  \omega  \cos (\pi  x)-\omega  \sin (\pi  x) (2 \pi  x \omega +(\omega +i \pi ) \sin (\pi  x))}{\pi ^2}+\frac{2 i \omega }{\pi }
\end{align*}
\begin{equation*}
    c_1=\frac{2 i \cos ^2\left(\frac{\pi  x}{2}\right) (2 \pi  x \omega +(2 \omega +i \pi ) \sin (\pi  x))}{\pi ^2} \hspace{3cm}  c_2=\frac{4 \cos ^4\left(\frac{\pi  x}{2}\right)}{\pi ^2}  
\end{equation*}
 
\noindent Now we can take the coefficient of $\omega^0$ in $c_0$ to get $c_{0,0}$, the coefficient of $\omega^1$ in $c_0$ to get $c_{1,0}$ and so on.

\subsection{Other methods: merits and demerits}

\noindent Apart from the methods discussed above, there are a plethora of analytic and numerical methods that can be employed to compute the quasinormal spectra. One can compare numerically obtained QNMs 
with exact results for the few effective potentials which are analytically solvable\cite{ferrari1984new,heidari2023investigation}. 
In some cases, the exactly solvable potentials are
used as approximations for the real potentials in black hole perturbations (for example, in the Schwarzschild, a matching may be done using the Poschl--Teller potential, as shown many years ago in \cite{ferrari1984new} ). The QNMs found following this route
however do not match accurately when compared with 
results using other numerical methods. In particular, the Re$(\omega)$ for the overtones show a mismatch.

\noindent The direct integration method is another method that we have used to verify the fundamental modes for our geometries. This method involves integrating the perturbation equation outward from one boundary, inward from another boundary and matching at a midpoint to ensure continuity \cite{chandrasekhar1975}. The matching generates a characteristic equation, the roots of which are the QNMs. This method is well suited for fundamental modes of any generic $V_{eff}$ but is not 
very useful for overtones \cite{pani2013advanced} (lecture notes).

\noindent The WKB method, a well-known semi-analytic, approximation tool,  can be extended to the study of QNMs. In both the asymptotically constant regions ($r_*\xrightarrow{}\pm\infty$), the solution is approximated as linear combinations of in-going and out-going waves \cite{schutz1985black} and matched with solutions 
obtained using the behaviour of the effective potential near its maximum. The first two corrections were provided in \cite{iyer1987black}, and \cite{konoplya2003gravitational} provided the higher-order corrections. This approach is particularly well-suited for effective potentials which have a single peak and approach a constant value at both horizons.

\noindent The continued fraction method is a highly accurate and widely used technique for computing both fundamental modes and overtones \cite{leaver1985}. Here, one transforms the Schrodinger-like equation using a proper ansatz. Next, the solution is expressed in a series form, and a finite-term recurrence relation is obtained, putting the series solution in the transformed equation. This relation is then solved to compute the $\omega_{QNM}$. However, this method is challenging to implement for complicated potentials.

\noindent Apart from the methods listed above, there are many more.
These include the Asymptotic Iteration Method (AIM) \cite{Cho2012},
Physics-Informed-Neural-Networks (PINNs) \cite{Cornell2022,Luna2024} etc.  A recent summary of available methods and codes may be found
in \cite{bhpqnm}.

\noindent In the results presented below, we have primarily used
the Prony fit (time domain) and the spectral method (SM) while finding the QNMs.

\section{QNMs for ultrastatic, generalized Bronnikov--Ellis wormhole} \label{Table}

\noindent We present here, in table form, the values of QNMs (accurate up to 3 decimal places) for the family of wormholes (with $b_0=1$). The time-domain profiles are computed with initial conditions: $\psi(0,v)=0.5$ and $\psi(u,0)=exp\left(-\frac{(u-10)^{2}}{10}\right)$ and the $\omega_{QNM}$ are extracted by fitting these time-domains from $t=0$ to $t=20$ with a stepsize of $1/60$. For the starred results, the fitting window is $t=0$ to $t=60$ (note $m=0,1,2$ cases). The amplitudes are also calculated using the Prony fit method. We have not kept the results obtained using the Prony fit method, which significantly vary for different fitting windows. \\
\noindent In this ultra-static spacetime, if $(\omega_R-i\omega_I)$ is a solution, then $(-\omega_R-i\omega_I)$ is also a solution, both having the same amplitude. In the tables below, we show only the solution with a positive real part and double its amplitude (accounting for the amplitude of both solutions).\\
\\

\begin{table}[H]
    \centering
    \footnotesize
    \setlength{\tabcolsep}{4pt}
    \renewcommand{\arraystretch}{1.2}
    \singlespacing
    \scalebox{1}{
    \begin{tabular}{|c|c||c|c|c||c|c|c|c|}
        \hline
         & & \multicolumn{3}{c||}{\textbf{Scalar Perturbation}} & \multicolumn{3}{c|}{\textbf{Axial Perturbation}} \\
        \hline
        $m$ & Mode & Spectral Method & Prony Fit & Amplitude & Spectral Method & Prony Fit & Amplitude \\
        \hline
        0 & 0 & $0.68137-0.61775i$ & $0.68402-0.61594i^*$ & $-$ & $-$ & $-$ & $-$ \\
          & 1 & $0.49683-2.17988i$ & $-$ & $-$ & $-$ & $-$ & $-$ \\
        \hline
        1 & 0 & $1.57271-0.52970i$ & $1.57274-0.52969i^*$ & $-$ & $-$ & $-$ & $-$ \\
          & 1 & $1.25957-1.70681i$ & $-$ & $-$ & $-$ & $-$ & $-$ \\
        \hline
        2 & 0 & $2.54666-0.51267i$ & $2.54925-0.51285i$ & $0.35745915$ & $1.73769-0.30514i$ & $1.73771-0.30513i^*$ & $0.2689666$ \\
          & 1 & $2.35300-1.57522i$ & $-$ & $-$ & $1.71239-1.02010i$ & $-$ & $-$ \\
          & 2 & $1.94641-2.75472i$ & $-$ & $-$ & $1.52495-2.03050i$ & $-$ & $-$ \\
        \hline
        3 & 0 & $3.53429-0.50692i$ & $3.53443-0.50686i$ & $0.30122677$ & $2.95241-0.40998i$ & $2.95253-0.41006i$ & $0.29811729$ \\
          & 1 & $3.39038-1.53568i$ & $-$ & $-$ & $2.86643-1.25767i$ & $-$ & $-$ \\
          & 2 & $3.09988-2.62324i$ & $3.09682-2.62481i$ & $0.16837004$ & $2.66056-2.19085i$ & $2.67518-2.17517i$ & $0.171151$ \\
        \hline
        4 & 0 & $4.52705-0.50433i$ & $4.52734-0.50424i$ & $0.26517902$ & $4.07625-0.44911i$ & $4.07646-0.44904i$ & $0.2640608$ \\
          & 1 & $4.41536-1.52340i$ & $-$ & $-$ & $3.99432-1.36285i$ & $-$ & $-$ \\
          & 2 & $4.19023-2.57208i$ & $4.19075-2.57242i$ & $0.1397532$ & $3.80100-2.30237i$ & $3.79816-2.30549i$ & $0.14061934$ \\
        \hline
        5 & 0 & $5.52231-0.50296i$ & $5.52282-0.50282i$ & $0.23964086$ & $5.15482-0.46725i$ & $5.15524-0.46714i$ & $0.239214$ \\
          & 1 & $5.42996-1.51789i$ & $5.41796-1.52258i$ & $2.009 \times 10^{-5}$ & $5.07306-1.40721i$ & $-$ & $-$ \\
          & 2 & $5.24705-2.54683i$ & $5.25123-2.54327i$ & $0.12190640$ & $4.90770-2.37002i$ & $4.90815-2.37157i$ & $0.12362226$ \\
        \hline
        6 & 0 & $6.51897-0.50214i$ & $6.51980-0.50195i$ & $0.22033499$ & $6.20882-0.47711i$ & $6.20954-0.47695i$ & $0.2201443$ \\
          & 1 & $6.44246-1.50964i$ & $6.44313-1.53221i$ & $1.886 \times 10^{-5}$ & $6.13729-1.43537i$ & $-$ & $-$ \\
          & 2 & $6.28646-2.53346i$ & $6.28819-2.53269i$ & $0.11151401$ & $5.99151-2.40927i$ & $5.99267-2.40806i$ & $0.1115693$ \\
        \hline
        7 & 0 & $7.51649-0.50162i$ & $7.51934-0.50106i$ & $0.20454891$ & $7.24818-0.48307i$ & $7.24931-0.48285i$ & $0.20497677$ \\
          & 1 & $7.44933-1.50895i$ & $7.45576-1.50817i$ & $2.309 \times 10^{-5}$ & $7.18438-1.45261i$ & $-$ & $-$ \\
          & 2 & $7.31541-2.52544i$ & $7.31985-2.52412i$ & $0.10156691$ & $7.05635-2.43256i$ & $7.05785-2.43166i$ & $0.10294975$ \\
        \hline
    \end{tabular}
    }
    \caption{\footnotesize $\omega_{\text{QNM}}$ values for generalised Bronnikov--Ellis wormhole ($n=2$) under scalar perturbations for $m=0$ - $7$ and axial perturbations for $m=2$ - $7$. [The QNM spectra for the $n=2$ case were published in \cite{kim2008wormhole,batic2024unified,blazquez2018scalar,alfaro2024quasinormal}]}
    \label{tab:qnm_scalar_axial_2}
\end{table}

\begin{table}[H]
    \centering
    \footnotesize
    \setlength{\tabcolsep}{4pt}
    \renewcommand{\arraystretch}{1.2}
    \singlespacing
    \scalebox{1}{
    \begin{tabular}{|c|c||c|c|c||c|c|c|}
        \hline
         & & \multicolumn{3}{c||}{\textbf{Scalar Perturbation}} & \multicolumn{3}{c|}{\textbf{Axial Perturbation}} \\
        \hline
        $m$ & Mode & Spectral Method & Prony Fit & Amplitude & Spectral Method & Prony Fit & Amplitude \\
        \hline
        0 & 0 & $0.83639-0.39420i$ & $0.83639-0.39418i^*$ & $-$ & $-$ & $-$ & $-$ \\
          & 1 & $1.78264-1.42784i$ & $-$ & $-$ & $-$ & $-$ & $-$ \\
        \hline
        1 & 0 & $1.68973-0.28405i$ & $1.68974-0.28404i^*$ & $-$ & $-$ & $-$ & $-$ \\
          & 1 & $2.28101-1.18068i$ & $-$ & $-$ & $-$ & $-$ & $-$ \\
        \hline
        2 & 0 & $2.64867-0.24369i$ & $2.64856-0.24379i$ & $0.30535863$ & $1.86212-0.27064i$ & $1.86214-0.27064i^*$ & $0.24995736$ \\
          & 1 & $3.09550-0.97920i$ & $-$ & $-$ & $1.97161-0.71811i$ & $-$ & $-$ \\
          & 2 & $3.73267-2.07179i$ & $3.72699-2.07264i$ & $0.22283351$ & $2.55089-1.37091i$ & $-$ & $-$ \\
        \hline
        3 & 0 & $3.60155-3.36888i$ & $3.62865-0.22143i$ & $0.24259909$ & $3.14398-0.31728i$ & $3.14405-0.31724i$ & $0.24589997$ \\
          & 1 & $4.01130-0.86237i$ & $-$ & $-$ & $3.18751-0.89811i$ & $-$ & $-$ \\
          & 2 & $4.52123-1.81580i$ & $4.52041-1.81625i$ & $0.18309362$ & $3.48149-1.49460i$ & $3.48082-1.49494i$ & $0.1574885$ \\
        \hline
        4 & 0 & $4.6164-0.20649i$ & $4.61663-0.20641i$ & $0.20468468$ & $4.28394-0.28792i$ & $4.28414-0.28787i$ & $0.21497073$ \\
          & 1 & $4.96092-0.78763i$ & $4.95844-0.78954i$ & $0.00019476$ & $4.40055-0.92088i$ & $-$ & $-$ \\
          & 2 & $5.40832-1.63672i$ & $5.41382-1.63803i$ & $0.15290777$ & $4.59293-1.60512i$ & $4.59135-1.60323i$ & $0.1200131$ \\
          & 3 & $5.94108-2.65940i$ & $-$ & $-$ & $4.98385-2.34634i$ & $-$ & $-$ \\
          & 4 &   $-$                    &   $-$                    &   $-$      & $5.60020-3.24404i$ & $5.57299-3.21739i$ & $0.12091509$ \\
        \hline
        5 & 0 & $5.60815-0.19523i$ & $5.60859-0.19513i$ & $0.1788866$ & $5.35662-0.25860i$ & $5.35702-0.25851i$ & $0.18764768$ \\
          & 1 & $5.92623-0.73455i$ & $5.92481-0.73471i$ & $0.00011546$ & $5.53163-0.86711i$ & $5.54394-0.87820i$ & $1.795\times10^{-5}$ \\
          & 2 & $6.33667-1.50966i$ & $6.33719-1.51452i$ & $0.13574164$ & $5.74869-1.60115i$ & $5.74805-1.60159i$ & $0.11547397$ \\
          & 3 & $6.81719-2.45107i$   & $-$                & $-$          & $5.91553-2.26906i$ & $-$                & $-$ \\
          & 4 &          $-$            &          $-$          &        $-$      & $6.46920-3.21880i$ & $6.41666-3.21904i$ & $0.0932583$ \\
        \hline
        6 & 0 & $6.60210-0.18623i$ & $6.60282-0.18610i$ & $0.15997014$ & $6.39998-0.23613i$ & $6.40069-0.23600i$ & $0.1668965$ \\
          & 1 & $6.90042-0.69409i$ & $6.90291-0.69404i$ & $1.259\times10^{-5}$ & $6.60250-0.80713i$ & $6.60367-0.81068i$ & $2.414\times10^{-5}$ \\
          & 2 & $7.28536-1.41469i$ & $7.28540-1.41346i$ & $0.11879929$ & $6.85764-1.53123i$ & $6.85660-1.52978i$ & $0.10977835$ \\
          & 3 & $7.72908-2.28317i$   & $-$                & $-$          & $7.21059-2.33977i$ & $-$                & $-$ \\
          & 4 & $8.24197-3.28172i$ & $8.25637-3.28542i$ & $0.09481262$ & $7.48045-3.19207i$ & $7.50483-3.1369i$ & $0.07558219$ \\
        \hline
        7 & 0 & $7.59739-0.17875i$ & $7.59853-0.17859i$ & $0.1454134$ & $7.42853-0.21893i$ & $7.42962-0.21877i$ & $0.15090914$ \\
          & 1 & $7.88016-0.66170i$ & $7.88124-0.66145i$ & $4.624\times10^{-5}$ & $7.64284-0.75603i$ & $7.64344-0.75595i$ & $6.13\times10^{-6}$ \\
          & 2 & $8.24579-1.34029i$ & $8.24539-1.33931i$ & $0.10773933$ & $7.91941-1.45107i$ & $7.91918-1.45030i$ & $0.10293865$ \\
          & 3 & $8.65746-2.16049i$ & $-$                & $-$          & $8.22111-2.20332i$ & $-$                & $-$ \\
          & 4 & $9.13470-3.08560i$ & $9.10375-3.06571i$ & $0.0848906$ & $8.54449-3.11196i$ & $8.54457-3.06192i$ & $0.06786849$ \\
        \hline
    \end{tabular}
    }
    \caption{\footnotesize $\omega_{\text{QNM}}$ values for generalised Bronnikov--Ellis wormhole ($n=4$) under scalar perturbations for $m=0$ - $7$ and axial perturbations for $m=2$ - $7$. [The fundamental modes for the $n>2$ cases were published in \cite{dutta2020revisiting} (scalar) and \cite{dutta2022novel} (axial)]}
    \label{tab:qnm_scalar_axial_4}
\end{table}

\begin{table}[H]
    \centering
    \footnotesize
    \setlength{\tabcolsep}{4pt}
    \renewcommand{\arraystretch}{1.2}
    \singlespacing
    \scalebox{1}{
    \begin{tabular}{|c|c||c|c|c||c|c|c|}
        \hline
         & & \multicolumn{3}{c||}{\textbf{Scalar Perturbation}} & \multicolumn{3}{c|}{\textbf{Axial Perturbation}} \\
        \hline
        $m$ & Mode & Spectral Method & Prony Fit & Amplitude & Spectral Method & Prony Fit & Amplitude \\
        \hline
        0 & 0 & $0.86815-0.34320i$ & $0.86815-0.34319i^*$ & $-$ & $-$ & $-$ & $-$ \\
          & 1 & $2.04851-1.11401i$ & $-$ & $-$ & $-$ & $-$ & $-$ \\
          & 2 & $3.38964-1.91968i$ & $3.39861-1.92833i^*$ & $-$ & $-$ & $-$ & $-$ \\
        \hline
        1 & 0 & $1.71264-0.22384i$ & $1.71265-0.22383i^*$ & $-$ & $-$ & $-$ & $-$ \\
          & 1 & $2.49623-0.91263i$ & $-$ & $-$ & $-$ & $-$ & $-$ \\
          & 2 & $3.64769-1.77155i$ & $3.64832-1.77127i^*$ & $-$ & $-$ & $-$ & $-$ \\
        \hline
        2 & 0 & $2.66518-0.17757i$ & $2.66523 - 0.17779i$ & $0.29315965$ & $1.89443-0.24761i$ & $1.89445-0.24761i^*$ & $0.2266797$ \\
          & 1 & $3.24903-0.73077i$ & $-$ & $-$ & $2.04700-0.61385i$ & $-$ & $-$ \\
          & 2 & $4.15868-1.54933i$ & $4.15833 - 1.54843i$ & $0.20476605$ & $2.78063-1.07127i$ & $2.77464-1.07131i^*$ & $0.30966731$ \\
          & 3 & $5.28869-2.41854i$ & $-$ & $-$ & $-$ & $-$ & $-$ \\
          & 4 & $6.52829-3.26724i$ & $6.53412 - 3.27928i$ & $0.40308849$ & $-$ & $-$ & $-$ \\
        \hline
        3 & 0 & $3.64016-0.15237i$ & $3.64024 - 0.15233i$ & $0.22714669$ & $3.18363-0.28391i$ & $3.18372-0.28388i$ & $0.23197406$ \\
          & 1 & $4.12259-0.61681i$ & $4.14739 - 0.63000i$ & $0.00015229$ & $3.25390-0.76629i$ & $-$ & $-$ \\
          & 2 & $4.86332-1.34256i$ & $4.86269 - 1.34213i$ & $0.18258948$ & $3.65647-1.17197i$ & $3.65640-1.17160i$ & $0.16536179$ \\
          & 3 & $5.81271-2.19639i$ & $-$ & $-$ & $4.52375-1.72969i$ & $-$ & $-$ \\
          & 4 & $6.94662-3.06045i$ & $6.95026 - 3.06342i$ & $0.12904273$ & $5.60501-2.47686i$ & $5.60406-2.47709i$ & $0.15103107$ \\
        \hline
        4 & 0 & $4.62419-0.13598i$ & $4.62440 - 0.13592i$ & $0.18790256$ & $4.32457-0.23666i$ & $4.32477-0.23661i$ & $0.20448213$ \\
          & 1 & $5.04395-0.54241i$ & $5.04408 - 0.54139i$ & $1.859\times10^{-5}$ & $4.47738-0.78946i$ & $-$ & $-$ \\
          & 2 & $5.67924-1.18324i$ & $5.67895 - 1.18265i$ & $0.15644501$ & $4.71713-1.30051i$ & $4.71834-1.29988i$ & $0.10481977$ \\
          & 3 & $6.51135-1.97082i$ & $-$ & $-$ & $5.30511-1.77828i$ & $-$ & $-$ \\
          & 4 & $7.49665-2.83332i$ & $7.49472 - 2.83864i$ & $0.11727697$ & $6.23010-2.39260i$ & $6.22995-2.38931i$ & $0.13343803$ \\
        \hline
        5 & 0 & $5.61290-0.12418i$ & $5.61332 - 0.12410i$ & $0.16150451$ & $5.39035-0.19645i$ & $5.39075-0.19637i$ & $0.17505116$ \\
          & 1 & $5.98915-0.48994i$ & $5.98930 - 0.48917i$ & $7.8\times10^{-6}$ & $5.62591-0.70862i$ & $5.62610-0.71015i$ & $4.315\times10^{-5}$ \\
          & 2 & $6.55411-1.06444i$ & $6.55411 - 1.06349i$ & $0.13585498$ & $5.87683-1.32855i$ & $5.87679-1.32565i$ & $0.09890309$ \\
          & 3 & $7.28065-1.79073i$ & $-$ & $-$ & $6.26018-1.85524i$ & $-$ & $-$ \\
          & 4 & $8.15775-2.61273i$ & $8.15569 - 2.60335i$ & $0.10450563$ & $6.96511-2.38395i$ & $6.9809-2.36109i$ & $0.10370454$ \\
        \hline
        6 & 0 & $6.60441-0.11513i$ & $6.60513 - 0.11501i$ & $0.14239417$ & $6.42671-0.16857i$ & $6.42738-0.16847i$ & $0.15284811$ \\
          & 1 & $6.94825-0.45059i$ & $6.94923 - 0.45040i$ & $0.00018714$ & $6.69168-0.62009i$ & $6.69217-0.62012i$ & $0.00011249$ \\
          & 2 & $7.46242-0.97399i$ & $7.46683 - 0.96882i$ & $0.1139611$ & $7.01628-1.24539i$ & $7.01625-1.24492i$ & $0.10544664$ \\
          & 3 & $8.11603-1.64203i$ & $-$ & $-$ & $7.35083-1.88535i$ & $-$ & $-$ \\
          & 4 & $8.90307-2.41503i$ & $8.79939 - 2.40900i$ & $0.06811125$ & $7.84927-2.43946i$ & $7.84363-2.43251i$ & $0.08016603$ \\
        \hline
        7 & 0 & $7.59771-0.10787i$ & $7.59880 - 0.10775i$ & $0.12789715$ & $7.44938-0.14872i$ & $7.45045-0.14859i$ & $0.13601879$ \\
          & 1 & $7.91626-0.41968i$ & $7.91704 - 0.41916i$ & $8.235\times10^{-5}$ & $7.71959-0.55112i$ & $7.72040-0.55064i$ & $3.859\times10^{-5}$ \\
          & 2 & $8.39152-0.90296i$ & $8.39220 - 0.90173i$ & $0.10758787$ & $8.08261-1.12967i$ & $8.08311-1.12881i$ & $0.10269501$ \\
          & 3 & $8.99190-1.52108i$ & $8.93263 - 1.51227i$ & $0.00374277$ & $8.46623-1.80756i$ & $-$ & $-$ \\
          & 4 & $9.70803-2.24495i$ & $9.69532 - 2.23123i$ & $0.09946345$ & $8.87902-2.45825i$ & $8.88319-2.49385i$ & $0.06576324$ \\
        \hline
    \end{tabular}
    }
    \caption{\footnotesize $\omega_{\text{QNM}}$ values for generalised Bronnikov--Ellis wormhole ($n=6$) under scalar perturbations for $m=0$ - $7$ and axial perturbations for $m=2$ - $7$. [The fundamental modes for the $n>2$ cases were published in \cite{dutta2020revisiting} (scalar) and \cite{dutta2022novel} (axial)]}
    \label{tab:qnm_scalar_axial_6}
\end{table}

\begin{table}[H]
    \centering
    \footnotesize
    \setlength{\tabcolsep}{4pt}
    \renewcommand{\arraystretch}{1.2}
    \singlespacing
    \scalebox{1}{
    \begin{tabular}{|c|c||c|c|c||c|c|c|}
        \hline
         & & \multicolumn{3}{c||}{\textbf{Scalar Perturbation}} & \multicolumn{3}{c|}{\textbf{Axial Perturbation}} \\
        \hline
        $m$ & Mode & Spectral Method & Prony Fit & Amplitude & Spectral Method & Prony Fit & Amplitude \\
      \hline
      0 & 0 & $0.87997-0.32519i$ & $0.87998-0.32518i^*$ & $-$ & $-$ & $-$ & $-$ \\
        & 1 & $2.14466-0.97491i$ & $-$ & $-$ & $-$ & $-$ & $-$ \\
        & 2 & $3.57108-1.59477i$ & $3.57054-1.59493i^*$ & $-$ & $-$ & $-$ & $-$ \\
      \hline
      1 & 0 & $1.72193-0.20105i$ & $1.72194-0.20104i^*$ & $-$ & $-$ & $-$ & $-$ \\
        & 1 & $2.57737-0.79070i$ & $-$ & $-$ & $-$ & $-$ & $-$ \\
        & 2 & $3.82365-1.47070i$ & $3.82587-1.46941i^*$ & $-$ & $-$ & $-$ & $-$ \\
      \hline
      2 & 0 & $2.67187-0.15181i$ & $2.67251-0.15238i$ & $0.29083766$ & $1.91003-0.23457i$ & $1.91005-0.23456i^*$ & $0.21776703$ \\
        & 1 & $3.30697-0.61908i$ & $-$                & $-$         & $2.08571-0.56896i$ & $-$                  & $-$         \\
        & 2 & $4.31495-1.28267i$ & $4.31427-1.28190i$ & $0.23210154$ & $2.87446-0.93747i$ & $2.87405-0.93741i^*$ & $0.29184553$ \\
        & 3 & $5.55107-1.96109i$  & $-$                & $-$         & $4.04950-1.50405i$ & $-$                  & $-$         \\
        & 4 & $6.88381-2.61925i$  & $6.88773-2.61918i$ & $0.25385346$ & $5.37019-2.11453i$ & $5.36933-2.11477i^*$ & $0.1670686$ \\
      \hline
      3 & 0 & $3.64456-0.12512i$ & $3.64471-0.12510i$ & $0.22202515$ & $3.19594-0.26645i$ & $3.19603-0.26648i$    & $0.22253281$ \\
        & 1 & $4.16276-0.50814i$ & $4.17354-0.50844i$ & $0.00037032$ & $3.28608-0.69780i$ & $-$                  & $-$         \\
        & 2 & $4.98966-1.10188i$ & $4.99007-1.10179i$ & $0.18332356$ & $3.73814-1.01798i$ & $3.73866-1.01791i$    & $0.17541431$ \\
        & 3 & $6.06575-1.77885i$ & $-$                & $-$         & $4.69073-1.44514i$ & $-$                  & $-$         \\
        & 4 & $7.29017-2.45688i$ & $7.29374-2.46700i$ & $0.13255104$ & $5.86562-2.02004i$ & $5.8667-2.02183i$     & $0.15024292$ \\
      \hline
      4 & 0 & $4.62667-0.10807i$ & $4.62688-0.10801i$ & $0.18186996$ & $4.33741-0.21472i$ & $4.33761-0.21467i$    & $0.19968836$ \\
        & 1 & $5.07164-0.43530i$ & $5.07146-0.43473i$ & $5.0481\times10^{-6}$ & $4.49808-0.72204i$ & $-$                  & $-$         \\
        & 2 & $5.77715-0.95771i$ & $5.77702-0.95720i$ & $0.15714858$ & $4.76792-1.13746i$ & $4.76742-1.13632i$    & $0.10174903$ \\
        & 3 & $6.71074-1.59843i$ & $-$                & $-$         & $5.44241-1.46518i$ & $-$                  & $-$         \\
        & 4 & $7.81676-2.27620i$ & $7.81867-2.27681i$ & $0.11845874$ & $6.47562-1.94993i$ & $6.47398-1.95101i$    & $0.13691266$ \\
      \hline
      5 & 0 & $5.61380-0.09608i$ & $5.61421-0.09601i$ & $0.15498730$ & $5.40063-0.17075i$ & $5.40103-0.17067i$    & $0.16992681$ \\
        & 1 & $6.00767-0.38426i$ & $6.00786-0.38394i$ & $1.7893\times10^{-5}$ & $5.65481-0.64034i$ & $-$                  & $-$         \\
        & 2 & $6.62899-0.84805i$ & $6.62886-0.84759i$ & $0.13681384$ & $5.90429-1.18229i$ & $5.90559-1.17820i$    & $0.08731672$ \\
        & 3 & $7.45181-1.43958i$ & $7.45249-1.39775i$ & $0.00057971$ & $6.34149-1.56146i$ & $-$                  & $-$         \\
        & 4 & $8.44520-2.09680i$ & $8.44251-2.10081i$ & $0.11003456$ & $7.17592-1.93339i$ & $7.18692-1.93325i$    & $0.11707173$ \\
      \hline
      6 & 0 & $6.60399-0.08711i$ & $6.60469-0.08702i$ & $0.13560445$ & $6.43410-0.14111i$ & $6.43478-0.14102i$    & $0.14713892$ \\
        & 1 & $6.95973-0.34643i$ & $6.96018-0.34614i$ & $1.3489\times10^{-5}$ & $6.72107-0.54020i$ & $6.72167-0.53992i$    & $8.009\times10^{-5}$ \\
        & 2 & $7.51913-0.76397i$ & $7.51916-0.76313i$ & $0.12066340$ & $7.05523-1.11725i$ & $7.05465-1.11561i$    & $0.09874688$ \\
        & 3 & $8.25731-1.30704i$ & $8.23586-1.33897i$ & $0.00457058$ & $7.39302-1.62858i$ & $-$                  & $-$         \\
        & 4 & $9.15431-1.93145i$ & $9.15519-1.93202i$ & $0.10141119$ & $7.98533-1.99226i$ & $7.98248-1.98655i$    & $0.08462117$ \\
      \hline
      7 & 0 & $7.59621-0.08009i$ & $7.59732-0.07999i$ & $0.12093846$ & $7.45434-0.12059i$ & $7.45540-0.12047i$    & $0.12992236$ \\
        & 1 & $7.92220-0.31715i$ & $7.92296-0.31677i$ & $3.2848\times10^{-5}$ & $7.74342-0.46352i$ & $7.74431-0.46321i$    & $1.434\times10^{-5}$ \\
        & 2 & $8.43378-0.69798i$ & $8.43393-0.69726i$ & $0.10853332$ & $8.13502-0.98974i$ & $8.13560-0.98861i$    & $0.10233494$ \\
        & 3 & $9.10647-1.19796i$ & $9.10056-1.19398i$ & $0.00094986$ & $8.51244-1.59506i$ & $-$                  & $-$         \\
        & 4 & $9.92416-1.78541i$ & $9.92248-1.77980i$ & $0.09058615$ & $8.94988-2.06737i$ & $8.95234-2.06370i$    & $0.06334415$ \\
      \hline
    \end{tabular}
    }
    \caption{\footnotesize $\omega_{\text{QNM}}$ values for generalised Bronnikov--Ellis wormhole ($n=8$) under scalar perturbations for $m=0$ - $7$ and axial perturbations for $m=2$ - $7$. [The fundamental modes for the $n>2$ cases were published in \cite{dutta2020revisiting} (scalar) and \cite{dutta2022novel} (axial)]}
    \label{tab:qnm_scalar_axial_8}
\end{table}

\end{document}